\documentclass[aps,a4paper,showpacs,superscriptaddress,twocolumn,floatfix]{revtex4}

\usepackage{hyperref}
\usepackage{graphicx}
\usepackage[sort&compress]{natbib} 
\usepackage[utf8]{inputenc}
\usepackage[english]{babel}
\usepackage{mathrsfs}
\usepackage{cancel}
\usepackage{amsmath}
\usepackage{amssymb}
\usepackage{bbm}
\usepackage{subfigure}
\usepackage{color}

\begin{document}
\title{Deep reinforcement learning for complex evaluation of one-loop diagrams in quantum field theory}
\date{\today}
\author{Andreas Windisch}
\email[Electronic address: ]{windisch@physics.wustl.edu}
\affiliation{Department of Physics, Washington University in St. Louis, MO 63130, USA}
\affiliation{Silicon Austria Labs GmbH, Inffeldgasse 25F, 8010 Graz, Austria}
\author{Thomas Gallien}
\email[Electronic address: ]{thomas.gallien@silicon-austria.com}
\affiliation{Silicon Austria Labs GmbH, Inffeldgasse 25F, 8010 Graz, Austria}
\author{Christopher Schwarzlm\"uller}
\email[Electronic address: ]{christopher.schwarzlmueller@silicon-austria.com}
\affiliation{Silicon Austria Labs GmbH, Inffeldgasse 25F, 8010 Graz, Austria}


\begin{abstract}
In this paper we present a novel technique based on deep reinforcement learning that allows for numerical 
analytic continuation of integrals that are often encountered in one-loop diagrams in quantum field theory. 
In order to extract certain quantities of two-point functions, such as spectral densities, mass poles or 
multi-particle thresholds, it is necessary to perform an analytic continuation of the correlator in question. 
At one-loop level in Euclidean space, this results in the necessity to deform the integration contour of the 
loop integral in the complex plane of the square of the loop momentum, in order to avoid non-analyticities 
in the integration plane. Using a toy model for which an exact solution is known, we train a reinforcement 
learning agent to perform the required contour deformations. Our study shows great promise for an agent 
to be deployed in iterative numerical approaches used to compute non-perturbative 2-point 
functions, such as the quark propagator Dyson-Schwinger equation, or more generally, 
Fredholm  equations of the second kind, in the complex domain. 
\end{abstract}
\maketitle

\section{\label{sec:intro}Introduction}
Studying two-point functions in the complex domain is a worthwhile endeavor, as valuable insights about 
the properties of the propagating degree of freedom become accessible.
Information about masses, multi-particle thresholds, possible composite nature, and even a sufficient criterion 
for removing the degree of freedom from the physical state space, can be derived once the analytic structure of 
the correlator is known. In the realm of non-perturbative Quantum Chromo Dynamics (QCD), the continuum approach of 
Dyson-Schwinger Equations (DESs) and Bethe-Salpeter Equations (BSEs) can be used to study hadron phenomenology, 
see e.~g.~\cite{Alkofer:2000wg,Fischer:2006ub,Roberts:2007jh,Sanchis-Alepuz:2015tha,Cloet:2013jya,
Eichmann:2016yit,Sanchis-Alepuz:2017jjd,Eichmann:2019vhk}, which also requires knowledge of the analytic 
properties of some of the quantities involved. Solving two-point functions with cubic vertices at one-loop level 
for complex momentum squares, however, is already a technically challenging task, as the integration contour of 
the radial variable associated with the loop momentum, expressed in hyperspherical coordinates in Euclidean 
space, cannot be maintained along the positive real half-axis once the external momentum square is allowed 
to assume complex values. Since there are two momenta involved, an external momentum, and a loop momentum, 
and because we consider cubic interactions, at least one of the propagators in the loop has to carry a mixed 
momentum. With an adequate choice of coordinates, the four-fold integral over the Euclidean loop momentum can thus 
be reduced to a two-fold integral over the radial component of the loop momentum, as well as over the angle between 
the external momentum and the loop momentum, see Appendix \ref{app:A}. 
As discussed on the basis of a very simple example in 
\cite{Windisch:2013dxa}, the angular integral then produces branch cuts in the complex plane of the radial 
integration variable, and one has to deform the contour in order to avoid the cut, as well as poles that might be 
present as well. Such strategies have been applied in various situations, see e.~g.~\cite{Maris:1995ns,
Strauss:2012dg,Windisch:2012zd,Windisch:2012sz,Weil:2017knt,Pawlowski:2017gxj,Williams:2018adr,
Eichmann:2019dts,Miramontes:2019mco}. While this task becomes very tedious already in a situation where the 
required deformations can be deduced from analyzing the analytic properties in the integration plane, 
in non-perturbative approaches, such as Dyson-Schwinger equations of two-point functions in the complex domain, 
it becomes extremely challenging, see e.~g.~\cite{Alkofer:2003jj,Strauss:2012dg,Windisch:2016iud} and 
references therein. These equations are such, that the function that is to be computed on the left hand side 
of the equation also appears in the integrand on its right hand side, which is the general form of a 
Fredholm integral equation of the second kind. The analytic properties of the integrand can thus only be determined 
for a given starting guess for the unknown function to be computed, but will change throughout 
the next iteration steps. A contour that was valid for the initial guess of the function would then have to 
be re-adjusted. The goal of this study is to provide a proof of principle that a deep reinforcement learning (DRL) agent
(see e.~g.~\cite{DRL} for a recent review and \cite{Sutton:2018aa} for a standard text book on the subject)
can be trained to conduct the contour deformations as needed. Such an agent could then be used in an iterative setting 
by deducing the contour deformation from observing the integration plane before each iteration step is conducted.
With this approach, an analytic continuation of Dyson-Schwinger Equations could become feasible.\\
The paper is organized as follows. In Section \ref{sec:model}, we review the toy model that we used to train the agent.
In Section \ref{sec:metho} we provide a short introduction to deep reinforcement learning, and 
Section \ref{sec:prep_env} addresses some of the prerequisites needed for training the agent on the toy model.
The numerical results are presented in Section \ref{sec:experiments}, and we
conclude in Section \ref{sec:summary}, where we specifically focus on how to further improve upon the agent's performance 
and also on the question of how to modify our approach such that it directly applicable 
to Dyson-Schwinger equations. Finally, Appendix \ref{app:A} gives an overview of our convention. 

\section{\label{sec:model}The toy model}
\subsection{Setting the stage}
The toy model we chose for this study features propagators of the Gribov type, the so-called $i$-particles \cite{Baulieu:2009ha}.
Apart from being physically interesting, there is another aspect that renders this model particularly useful
for our cause. The correlator that we are interested in has an exact solution, and has been used in a study
on numerical contour deformations \cite{Windisch:2012zd} that reproduced the analytic result perfectly.
This allows us to compare the deformations provided by the trained agent with the ones that have been obtained 
analytically in \cite{Windisch:2012zd}, where the criterion of similarity is qualitative, as we discuss in
this paper. With that, we can deduce that we could solve the integral successfully without 
the necessity of actually evaluating it.

The starting point for our study is given by equation (32) in \cite{Baulieu:2009ha}, which is an expression for a
correlator that has also been the main focus of \cite{Windisch:2012zd}. In $D$ dimensions, the correlator can be 
expressed as

\begin{equation}
\label{loop_diagram}
\mathscr{G}(p^2)=\int\frac{d^D q}{(2\pi)^4}\frac{1}{(p-q)^2-i\sqrt{2}\theta^2}\frac{1}{q^2+i\sqrt{2}\theta^2},
\end{equation} 
with an external 4-momentum $p$ and loop 4-momentum $q$. The mass parameter $\theta$ is given
through the relation

\begin{equation}
2\sqrt{2}\theta^2\equiv 1.
\end{equation}
$\mathscr{G}(p^2)$ is the Euclidean momentum space operator corresponding to the correlator of squared Yang-Mills field-strength tensors,

\begin{equation}
\label{fsquare}
\langle F^2(x) F^2(0)\rangle = \int\frac{d^Dp}{(2\pi)^D}\exp\{i p\cdot x\}\mathscr{G}(p^2).
\end{equation}

The Yang-Mills field-strength tensor is defined by
\begin{equation}
F_{\mu\nu}^a=\partial_\mu A_\nu^a - \partial_\nu A_\mu^a + gf^{abc}A_\mu^bA_\nu^c,
\end{equation}
with latin indices corresponding to color, greek indices are Lorentz indices, $g$ is the coupling and $f^{abc}$ are the structure constants. Contracting all indices yields the square of the tensor that enters the correlator (\ref{fsquare}),
\begin{equation}
F^2 = F_{\mu\nu}^a F_{\mu\nu}^a.
\end{equation}
Equation (\ref{loop_diagram}) is the expression the we would like to solve in $d=4$ dimensions. 
In \cite{Baulieu:2009ha}, an exact solution to this integral is presented (after it has been regularized), which reads

\begin{equation}
\label{orig_sol}
\mathscr{G}_{sub}(x)=\frac{1}{16\pi^2}\left(1-\frac{\pi}{2x}+\frac{\sqrt{1-x^2}}{x}\arccos(x)\right),
\end{equation}
where the subscript $sub$ indicates, that the original integral has been regularized in the spirit of Bogoliubov, Parasiuk, Hepp and Zimmermann  (BPHZ) \cite{Bogoliubov:1957gp,Bogolyubov:1980nc,Hepp:1966eg,Zimmermann:1969jj}. The detailed calculation is described in \cite{Baulieu:2009ha}. Note, that the $x$ in equation (\ref{orig_sol}) is not a spatial coordinate, but corresponds to the complex square of the external loop momentum, that is, $x\equiv p^2 \in\mathbb{C}$.
After switching to hyperspherical coordinates (see Appendix \ref{app:A}), regularizing the integral using the BPHZ scheme, as well as integrating the two trivial angles of the 4-dimensional hypersphere, the loop integral, equation (\ref{fsquare}), becomes

\begin{eqnarray}
\label{integral}
\mathscr{G}_{sub,rescaled}(x)=\frac{2}{\pi}\int_0^\infty dy\frac{y}{y^2+\frac{1}{4}}\\
\times\int_{-1}^{+1}dz\sqrt{1-z^2}\frac{-x+2\sqrt{x}\sqrt{y}z}{\left(x+y-2\sqrt{x}\sqrt{y}z-\frac{i}{2} \right)},\nonumber
\end{eqnarray} 
where the additional subscript $rescaled$ indicates that the loop integral (\ref{fsquare}) has been multiplied by a factor of $16\pi^2$ to eliminate the same factor in the denominator of the solution (\ref{orig_sol}). As before, $x$ is the complex external momentum square, and $y$ is the complex square of the loop momentum $q$, that is, $y\equiv q^2\in\mathbbm{C}$. $z$ is the cosine of the angle between the external and the loop momentum, $z\equiv\cos\theta$. For details as of how to arrive at equation (\ref{integral}) when starting with equation (\ref{fsquare}), see \cite{Windisch:2012zd}.

\subsection{The goal}
Now that we have revisited all the required expressions, let us state the goal of this study.
In \cite{Windisch:2012zd} it was shown, that the results of the integral equation for the $F^2$ correlator 
presented (and analytically solved) in \cite{Baulieu:2009ha}
can be reproduced numerically by applying suitable contour deformations.
The requirements for successful contours are:
\begin{itemize}
\item They avoid the branch cut in the integration plane.
\item They satisfy continuous deformability with respect to the original contour along the positive real half-axis
in a setting where the branch cut is absent and the poles are present.
\end{itemize}
The second requirement ensures, that no residue of any pole is picked up by the contour.
We thus consider the agent to be successful, if it is able to produce contours that satisfy these conditions. 
\section{\label{sec:metho}Methodology}

The goal is to find a valid integration contour in the complex plane of the radial integration
variable $y$ for any given external momentum square $x$.
In our approach we achieve this goal by reformulating the problem such that it becomes
applicable to deep reinforcement learning. More specifically, we define the task of contour
deformation as an episodic game, where a reinforcement learning agent is trained to escape a
maze in the presence of traps, trying to reach a dedicated target location.
The analogy to the original problem is as follows: The branch cut essentially defines a maze in a 
continuous space and cannot be crossed by the agent, while the target location is given by an $\epsilon$-area
around the cut-off on the positive real half-axis. Since the agent starts each episode at the origin of the
complex plane, the path taken to the target location corresponds then to the desired contour that solves the problem.
The traps are basically represented by the poles, since, if the agents picks up a residue along 
its way to the target location, the game is considered to be lost, regardless of whether the agent 
reaches the target location.

\subsection{Deep reinforcement learning}

Deep reinforcement learning describes a class of goal-orientated machine learning algorithms
taking advantage of powerful function approximators in the context of deep learning \cite{DRL,Sutton:2018aa}.
Unlike supervised or unsupervised machine learning, these algorithms do not require a dedicated
set of training data, since they are designed to learn from experience by interacting with
their environment. The key ingredient thereby is a scalar reward signal, which essentially
reinforces the learning entity to pick the desired actions.
It is important to note that the reward signal should not be mistaken as an error signal 
in a supervised setting, since it can, but doesn't need to, represent an error signal, nor is
it required to be differentiable. One can encounter situations where the reward
signal is sparse and rarely available, like e.~g.~in case of a game where the information about
failure or success is provided only at the end of an episode, which poses additional challenges.

\subsection{Fundamentals and Definitions}
The fundamental working principle of reinforcement learning algorithms is based upon the
concept of Markov decision processes (MDPs) and involves two entities, an \textit{agent}, and
an \textit{environment}~\cite{Sutton:2018aa}. Given a state $s_t \in \mathcal{S}$ at time step $t$, the agent
interacts with the environment by picking an action $a_t \in \mathcal{A}(s_t)$ according
to a policy $\pi(a_t|s_t)$, where $\mathcal{S}$ and $\mathcal{A}(s_t)$ denote the state- 
and action space, respectively. The policy is considered to be stochastic in general,
hence the agent samples the action from a conditional distribution. In that sense, we
treat a deterministic policy $\mu(s_t)$ as a special case $\pi(a_t|s_t)=\delta(a_t - \mu(s_t))$,
where $\delta(.)$ denotes the Dirac distribution. The environment responds to the action
$a_t$ by setting the consecutive next state $s_{t+1}$ with probability 
$\text{Pr}\{S_{t+1}=s_{t+1} |s_t, a_t\}$
\footnote{In our notation we use lowercase characters for the realization of the the corresponding random variable, 
which we denote in upper case} and gives rise to a 
reward $r_{t+1} \in \mathcal{R}$. Since the reward is considered to be stochastic as 
well, the dynamics of the MDP is entirely determined by the probability 
$\text{Pr}\{S_{t+1}=s_{t+1}, R_{t+1}=r_{t+1} | s_t, a_t\}$.

The goal in reinforcement learning is to find a policy $\pi(.)$ such that the
cumulative expected reward is maximized. The definition of the return
$G_t = \sum_{k=0}^\infty \gamma^k r_{t+k+1}$ provides a performance measure, where the
parameter $\gamma$, $0 < \gamma \leq 1$, denotes the discount rate and essentially
controls the impact of future states on the agent's decisions. Furthermore, we introduce the state-value
function $v_\pi(s) = \mathbb{E}_\pi\left\{G_t|s_t\right\}$ and the action-value function
$q_\pi(s,a) = \mathbb{E}_\pi\left\{G_t|s_t, a_t\right\}$,
which are both maximized by an optimal policy $\pi_\ast(.)$ 
for all $s\in \mathcal{S}$ and $a\in \mathcal{A}(s)$, hence 
$v_\ast(s) = \max_{\pi}v_\pi(s)$ and $q_\ast(s,a) = \max_{\pi}q_\pi(s,a)$. In case of a
finite MDP, $\mathcal{S}$, $\mathcal{A}$ and $\mathcal{R}$ are finite sets and an optimal
policy $\pi_\ast(.)$ can be derived by solving the Bellman optimality equation either for
the state-value function
\begin{equation}
v_\ast(s_t) = \max_{a}\mathbb{E}\left\{R_{t+1} + \gamma v_\ast(S_{t+1})|s_t,a_t\right\} ,\,
\end{equation}
or the action-value function
\begin{equation}
q_\ast(s_t,a_t) = \mathbb{E}\left\{R_{t+1} + \max_{a'}\gamma q_\ast(S_{t+1},a')|s_t, a_t\right\} .\,
\end{equation}

\subsection{Approximate Methods}
Finding optimal policies by solving Bellman's optimality equations requires a model of the 
environment in form of a discrete state transition distribution $p(s_{t+1}|s_t, a_t)$ and 
is thus not feasible for the vast majority of applications. Hence, reinforcement learning 
algorithms deal with approximate solutions and rely on a random component in order 
 to explore the environment. Always picking the action that maximizes
a value function estimate (greedy action) prevents the agent from exploring the state space. 
Consequently, approximate methods pick the greedy action only with a certain probability.
This probability is controlled by a parameter and is increased with increasing confidence 
in the value function estimate.

Provided that $\mathcal{S}$ defines a discrete state space, Monte Carlo methods produce
unbiased estimates of value functions by observing entire sequences of state-action-reward
tuples, in order to calculate the return $G_t$. 
Although unbiasedness is a striking argument, Monte Carlo methods suffer from the usual drawbacks
like e.~g.~high variance estimates and tremendous computational costs, since updates are solely 
performed after an entire episode has been observed. In contrast, temporal difference (TD)
methods update the value function by bootstrapping, which leads in general to much faster
convergence, but biased estimates. 

So far we assumed a discrete state space $\mathcal{S}$, which essentially limits the applicability
of Monte Carlo and TD methods to a specific set of applications. Moreover, we assumed
that the state space is entirely observable by the learning entity. However, this is often
not the case, since environmental observations are possibly continuous and only functionally 
related to the underlying states. Hence, the learning entity must be capable to generalize and to 
infer the relevant information from the observations. This is where function approximators, and
thus deep neural networks, come into play.

Value-based approaches utilize neural networks to approximate either state-value or action-value 
functions in order to derive a well-performing policy (see e.~g.~\cite{Mnih:2015aa}).
However, this approach is impractical for problems dealing with a continuous action space, since
picking the greedy action involves a $\max$ operation with respect to all $a\in\mathcal{A}(s)$.
Consequently, this would require to solve an optimization problem at every iteration. In contrast,
policy-based methods directly target the policy and are thus applicable to continuous control problems.

\subsection{Policy Gradient Methods}

Policy gradient methods aim to maximize the expected return (or a related 
value-based measure) by updating a parameterized policy $\pi_\theta$ by
means of gradient ascent~\cite{Sutton:1999aa}. In case the advantage function 
$A_\pi(s,a) = q_\pi(s,a) - v_\pi(s)$ is used for assessing the policy's performance,
the gradient of the loss for updating the policy network is given by 
\begin{equation}
 g = \mathbb{E}_{\tau\sim\pi_\theta}
 \left\{\sum_{t=0}^\infty\nabla\theta\log\pi_\theta(a_t|s_t)A_{\pi_\theta}(s_t,a_t)\right\} ,\,
\end{equation}
where $\tau = (s_0,a_0, \hdots, s_H,a_H,s_{H+1})$ denotes a trajectory of state and action
values up to horizon $H$ generated by sampling actions from the policy network $\pi_\theta$. 
Since deep learning frameworks rely on stochastic gradient like optimization algorithms,
we can define the policy gradient loss per training iteration as
\begin{equation}
L^{\text{PG}}(\theta) = \mathbb{\hat{E}}
 \left\{\log\pi_\theta(a_t|s_t)\hat{A_t}\right\} ,\,
\end{equation}
where $\mathbb{\hat{E}}$ denotes the empirical mean over a batch of samples and $\hat{A_t}$
is an estimate of the advantage function at time step $t$.

However, vanilla policy gradient methods use only first-order derivatives for updating the
policy network, and thus, the step size plays a crucial role. If chosen too small,
the agent learns too slowly in order to produce a well-performing policy, and if chosen
too big, the agent might pick an action which is very far from the greedy one.
Since the loss function is non-convex, there is a high risk that the update step overshoots. 
Hence, the agent proverbially falls off a reward cliff and the performance drops,
possibly without any prospect of recovery.
Another issue is the poor sample efficiency of 
vanilla policy gradient methods, since an entire trajectory is used to perform one single update.
Consequently, a lot of interaction with the environment is required in order to train the policy
network.

\subsection{Proximal Policy Optimization}
Trust region policy optimization (TRPO)~\cite{Schulman:2015aa} mitigates the shortcomings of 
vanilla policy gradient methods. On the one hand, trust region optimization algorithms
are much more robust when dealing with non-convex problems. These
methods first set an `area of interest' before determining the direction of the consecutive
optimization step. On the other hand, TRPO also increases the sample efficiency per trajectory, since
samples originating from a previous policy can be used to update the policy network by means of an
importance sampling scheme. Hence, the optimization (surrogate) objective for the TRPO agent is
\begin{equation}
\underset{\theta}{\max}\ \mathbb{\hat{E}}\left\{\frac{\pi_\theta(a_t|s_t)}{\pi_{\theta_\text{old}}(a_t|s_t)}\hat{A}_t\right\} 
\end{equation}
subject to
\begin{equation}
\mathbb{\hat{E}}\left\{D_\text{KL}\left(\pi_{\theta_\text{old}}(.|s_t) || \pi_\theta(.|s_t)\right)\right\} \leq  \delta ,\,
\end{equation}
where $D_\text{KL}(.)$ is the Kullback-Leibler divergence and $\delta$ is a hyper-parameter 
that defines the size of the trust region. We can convert this objective in an
unconstrained optimization problem by using a penalty term instead of the constraint. Nevertheless,
the need for calculating the Kullback-Leibler divergence still persists for every policy, 
which is computationally expensive.

Proximial Policy Optimization~\cite{Schulman:2017aa} (PPO) utilizes the key concepts of TRPO, but 
avoids to explicitly calculate the Kullback-Leibler divergence. The idea is surprisingly simple:
Let $\eta_t$ define the policy ratio used in the surrogate objective before,
\begin{equation}
 \eta_t(\theta) = \frac{\pi_\theta(a_t|s_t)}{\pi_{\theta_\text{old}}(a_t|s_t)}.
\end{equation}
Then the loss for each policy update is given by
\begin{equation}
L^{\text{PPO}}(\theta) = \mathbb{\hat{E}}
 \left\{\min\left(\eta_t\hat{A_t}, \text{clip}\left(\eta_t, 1-\epsilon, 1+\epsilon\right)\hat{A_t}\right)\right\} ,\,
\end{equation}
where $\epsilon$ is a hyper-parameter and usually chosen to be $0.1 \leq \epsilon \leq 0.3$.
Hence, clipping the policy ratio to values near $1$ has a similar effect as using a KL-constraint
in constraint optimization.

\section{\label{sec:prep_env}Constructing the environment}
Our PPO RL agent is trained by immersing it to a virtual environment in which the agent can act. 
The environment, in this case, is the complex $y$-plane, that is, the complex plane of the radial integration variable. 
The training goal of the agent is to find a path that connects the origin with the ultraviolet cutoff on the 
positive real axis for any given complex value of $x$, while avoiding the branch cut and the poles in the plane. 
Before the environment can be implemented, a thorough analysis of the integrand and the resulting cut structure is 
in order, as the information obtained from this analysis will be used directly to guide the agent throughout its 
learning phase. As a first step, we will assume that the information about the branch cut, as well as about the poles, 
is available in an analytic form. This is not a big restriction, as, even in self-energy expressions for DSEs, 
this information can be (at least partially) obtained exactly. In Section \ref{sec:summary} we furthermore discuss 
strategies as of how to pre-process data on the analytic properties of the integrand that is solely available numerically,
in order to make the strategy developed in this paper applicable to those cases as well. 
For now, we will direct our focus on extracting the relevant information about the analytic properties of the regularized 
integrand in a scenario where all information is accessible.


\subsection{\label{cut_structure}Branch cut structure}
\begin{figure}[h]
\includegraphics[width=8cm]{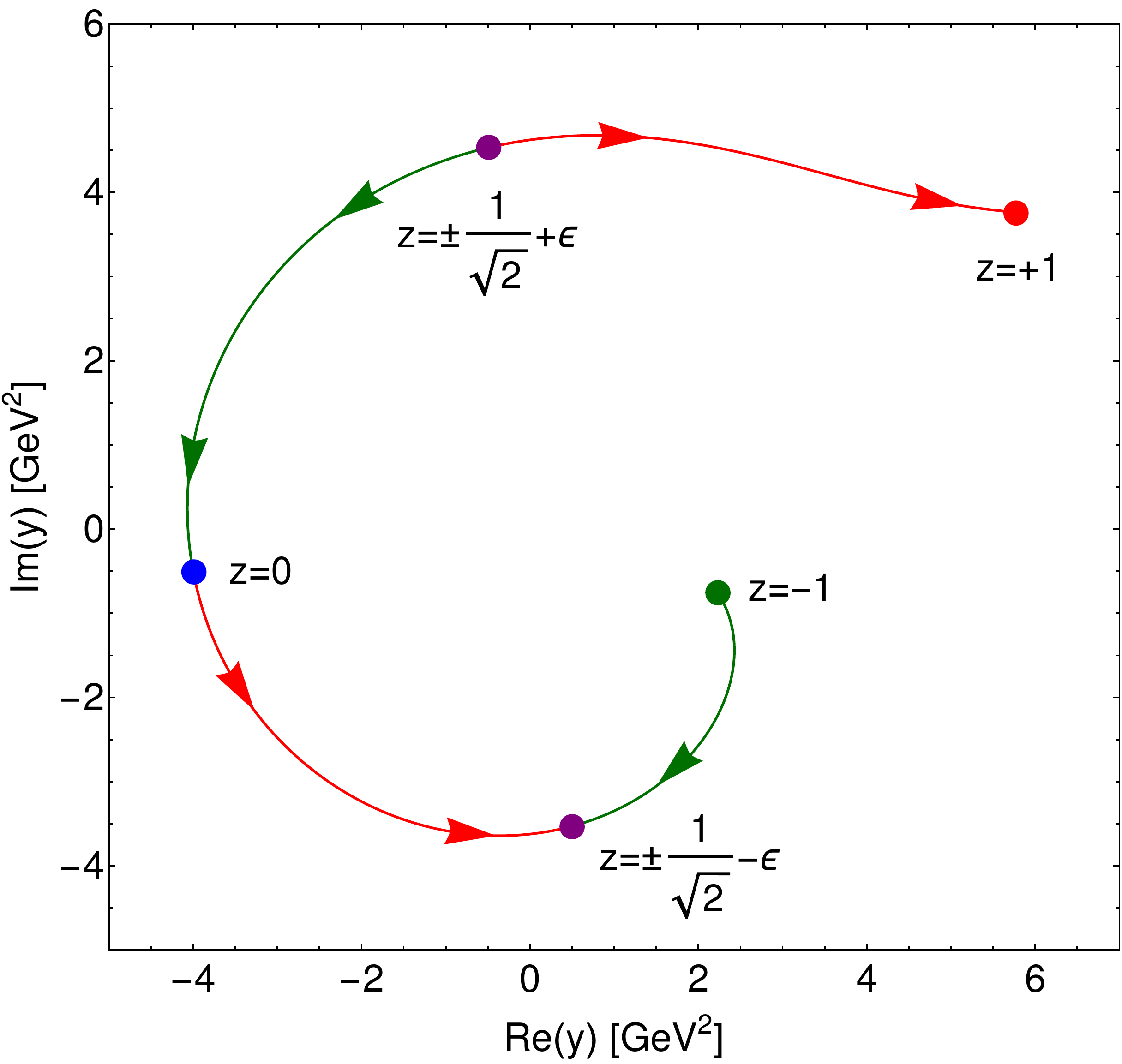}
\caption{The parametrization of the branch cut through the angular variable $z$ turns out to be discontinuous 
for some values of $x$. The image shows an example of such a discontinuity, here for $x=4+i$. For this value 
of $x$, the branch cut (\ref{cut_parametrization}) acquires the shape shown in the image. Starting at $z=-1$, 
with increasing value for $z$, the point $-\frac{1}{\sqrt{2}}-\epsilon$ is approached. At $z=-\frac{1}{\sqrt{2}}$, 
the parametrization jumps to the point $-\frac{1}{\sqrt{2}}+\epsilon$, and then approaches the first point from the 
other side, until it finally jumps back to the second point and runs towards $z=+1$, now tracing out the entire branch cut.}
\label{fig:cut_disc}
\end{figure}

In this Section we will analyze the integrand thoroughly, such that we can set up a training environment 
for the RL agent based on this information. The task of the agent will be to find valid integral contours for 
the $y$-integral in (\ref{integral}). The first term produces two poles, located at $y_{p_{1,2}}=\pm\frac{i}{2}$. 
Furthermore, a branch cut appears in the complex $y$-plane, induced by the angular integral over the variable $z$. 
As outlined in \cite{Windisch:2013dxa,Windisch:2012zd}, we have to solve for the zeros of the denominator while keeping 
$x$ fixed, and while varying $z$ from -1 to +1. This yields two parametrizations,

\begin{equation}
\label{cut_parametrization}
\xi_\pm(x,z)=\left(\sqrt{x}z\pm\sqrt{-x(1-z^2)+\frac{i}{2}}\right)^2,
\end{equation}
and it suffices to consider just one of these two congruent parametrizations.
There is, however, a problem that we have to address at this point. It turns out, that for some values of $x$, 
the parametrization of the cut becomes discontinuous in $z$, as explained in Figure \ref{fig:cut_disc}.

In order to being able to use the information about the non-analyticities in the complex $y$-plane to
construct the environment for the reinforcement learning agent in a convenient way, this discontinuity 
must be addressed. The occurrence of the discontinuity in the parametrization comes from the fact that the
radicand of the second square root in the branch cut parametrization (\ref{cut_parametrization}) changes its sign 
(i.~e.~the result lies on the other Riemann sheet of the square root as the one we want it to be on). 
The parametrization can be rendered continuous by insertion of the proper sign in the appropriate region, 
that can be determined, and expressed, analytically. However, in our implementation we chose a numerical approach to resolve 
this issue as follows. 

One key ingredient for this environment is to determine whether a given contour, comprised of a sequence
of connected line segments, intersects the branch cut at any point. We thus sample points from the
(possibly for a given $x$, discontinuous) branch cut parametrization. Then we start at the point $z=-1$ 
and choose the point among the sampled values with the shortest distance to the starting point.
Since the points on the branch cut are not distributed uniformly as $z$ is varied in fixed increments,
we introduced a maximally allowed distance between any two neighboring points. This is important to 
have control over the accuracy of the collision detection mechanism. In case we find this maximally 
allowed distance to be violated, we simply increase the number of points until 
the desired resolution is achieved. Repeating this procedure produces an ordered sequence of sampled points 
that we can then use in the branch cut collision detection.


\subsection{\label{sec:vector_env}Vectorized observations}

We provide the agent with relevant information about its environment as described 
in Table \ref{tab:obs}.

\begin{table}
\begin{tabular}{|c|c|c|}

\hline
\textbf{property}       & \textbf{domain}        &   \textbf{description} \\
\hline

$ |x| $        & $\mathbb{R}$  &  modulus of ext. momentum square \\
\hline

$ \arg x $     & $[-\pi,\pi]$    &  argument of ext. momentum square \\
\hline

$ \Lambda^2 $  & $\mathbb{R}$  &  cut-off on real axis (contour endpoint)\\
\hline

$ y_{p_1} $    & $\mathbb{C}$  &  location of first pole\\
\hline

$ y_{p_2} $    & $\mathbb{C}$  &  location of second pole\\
\hline

$ \xi_{i} $    & $\mathbb{C}$  &  starting point of branch cut\\
\hline

$ \xi_{f} $    & $\mathbb{C}$  &  endpoint of branch cut\\
\hline

$ y_A       $    & $\mathbb{C}$  &  position of agent in integration plane\\
\hline

$ \omega       $    & $\{T,F\}$  & agent left cut structure \\
\hline

\end{tabular}
\caption{Observables available for the RL agent.}
\label{tab:obs}
\end{table}

\begin{figure}[h]
\includegraphics[width=8cm]{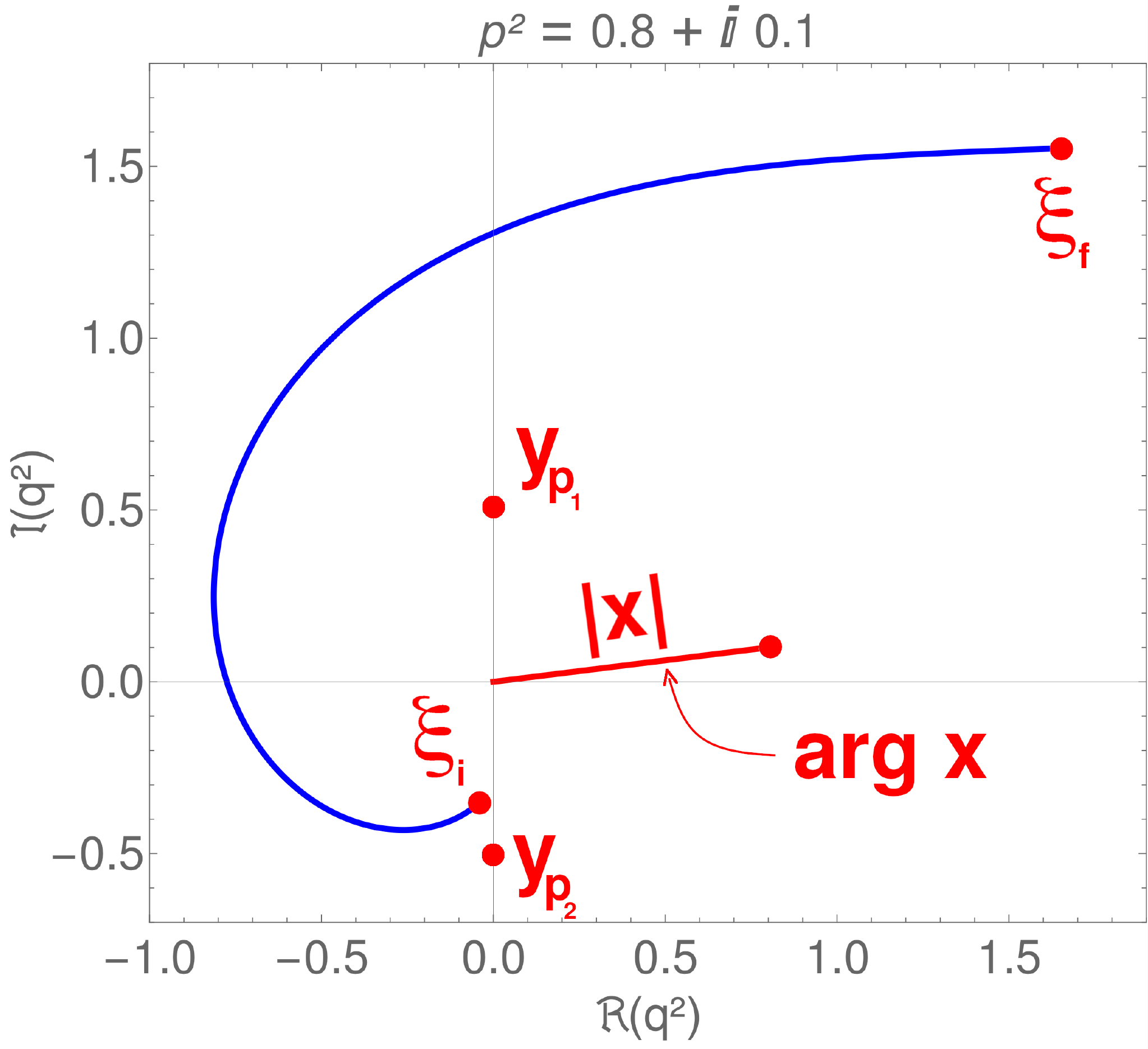}
\caption{Properties observed by the reinforcement learning agent.}
\label{fig:obs}
\end{figure}

The modulus $|x|$ of the external momentum is related to the extent of the branch cut structure, and
the argument $\arg{x}$ determines the direction of the opening of the cut structure.
In addition, we provide the starting- and endpoint ($\xi_i,\xi_f)$ of the branch cut structure, which always
lie on different sides of the line defined by $\arg{x}$. The cutoff $\Lambda^2$ is the 
point on the positive real axis that the agent has to reach, and $y_{p_i}$ denotes the location of two
complex conjugate poles. The position of the agent in the plane is given by $y_A$, and we also
provide a boolean value $\omega$ that indicates whether the agent has left the cut structure or not,
which is determined by looking for an intersection between the agent's path and the line connecting
$\xi_i$ with $\xi_f$. Based on this information, the agent is able to learn a policy that produces suitable contours.
Figure \ref{fig:obs} shows a depiction of the observed quantities.

\subsection{\label{sec:act}Actions}
Standing at any point in the complex plane, the agent chooses a direction and a distance for its next step. 
Its action space is thus 2-dimensional and continuous. The neural network responsible
for the agent's actions has two output neurons, each of which is activated by a hyperbolic
tangent. Since the hyperbolic tangent assumes values within the range of $[-1,1]$, the
actions of the agent are determined as follows. The length of the next step is
computed by adding 1 to the output of the associated output neuron, which maps the output
to the interval $[0,2]$. This quantity is then multiplied with $\frac{\Lambda^2}{2}$, where $\Lambda^2$ is 
not the numerical cutoff of the integral, but the point $y=10$ GeV$^2$, as discussed below. The
agent can thus choose any value between $0$ and $10$ Gev$^2$ for its stride in the complex plane.
Apart from the distance, the agent also has to choose a direction for its next step. This can 
be achieved by simply multiplying the output of the associated output neuron with $\pi$, which
can then be interpreted as the complex argument that determines the direction of the next step.

\subsection{\label{sec:col_det}Pole- and collision detection}
With every step taken, we then have to check whether a collision with the branch cut occurred or not.
The branch cut is approximated by line segments spanned by the ordered list of 
points described above. Detecting a collision thus boils down to just checking for an intersection
between the line segment from the previous point to the new point the agent stepped to, 
with all the line segments that comprise the approximated branch cut. 
If a collision is detected, the environment throws the agent back to 
where it was before it chose to take that step, and it can come up with a new suggestion.

The ultraviolet cutoff of the radial integral that has been used for the calculations in 
\cite{Windisch:2012zd} was $\Lambda^2= 10^4$ GeV$^2$. A successful candidate for a contour thus starts
at the origin $y=0$ and ends at $y=10^4$. However, we are only interested in evaluating the correlator
in a complex region around the origin that is at least three orders of magnitudes smaller than this cutoff.
As the typical scale of the non-analyticities arising in the integration plane is set by $|x|$, and because
this value is at most of the order $1$, the integrand is analytic along the real axis for values of 
$x\approx 10$ GeV$^2$ or greater. We thus set the point the agent has to reach to $y=10$ GeV$^2$. The rest of the contour
can then be added as the line connecting this point with the actual ultraviolet cutoff of the integral.

There is also another reason why this particular setup is preferable. If this approach is applied to
non-perturbative settings -- such as Dyson-Schwinger equations -- in the future, one will have to restrict
the area in the complex plane for which the equation is to be evaluated to a small region around the origin.
Since it lies within the nature of DSEs that the analytic properties of the integrand are not fully 
accessible, one will have to introduce a boundary, along which one closes the contour to a point on the 
positive real half axis not too far off from the origin. For this to be justifiable, one would have to 
look for evidence that suggests that the errors induced by this restriction have little impact on the result.
For example, it is imaginable that the discontinuities of the branch cuts become smaller with increasing $|x|$, 
which would justify this procedure and allow for a direct mapping from a DSE setting to the environment at hand.

In our model, we constrain the agent to the square $x\in [-1.5\Lambda^2,1.5\Lambda^2]$ GeV$^2 \times i[-1.5\Lambda^2,1.5\Lambda^2]$ GeV$^2$,
and we treat any step that takes the agent out of this region similar to it trying to cross a branch cut.
Since the analytic properties of the toy model are accessible exactly, we would not have to impose this 
restriction. However, it is beneficial in at least two ways. First, we will have to impose such a restriction 
in a future setting for DSEs, so we will be able to directly map the DSE setup on the one developed here, 
as discussed in Section \ref{sec:summary} below. Second, it also helps the agent to stick around the origin 
and thus close to the endpoint, which helps it to find its target more quickly.
  
Once the agent successfully enters a small region around the endpoint $y=10$ GeV$^2$, we have to check whether the path
suggested by the agent is continuously deformable into the original path along the positive real
half-axis if we neglect the presence of the branch cut, but maintain the poles.  
In order to determine whether this is the case, we (numerically) integrate a known function
that has poles with residue 1 at the same position as the integrand, but no branch cut, along
the suggested contour. If we then add the value of the integral along the real axis from $y=10$ GeV$^2$ to $y=0$ GeV$^2$, evaluated on
the same known function, we can deduce whether the path is continuously deformable in the sense discussed 
above by exploiting Cauchy's integral formula.

\subsection{\label{sec:rew}Reward function}
The reward function is a crucial quantity, as it allows us to directly encourage or discourage the agent
to take certain actions in certain situations. We punish the agent for failing to produce a
valid contour and for bumping into the branch cut (or the boundary) by providing it with a negative reward. We furthermore 
associate a small punishment for every step taken, which encourages the agent to take as few steps as 
possible to reach the goal. The agent receives a positive reward if it manages to reach the goal along a 
valid contour. We furthermore introduce a small reward or punishment depending on whether the last step
took the agent closer to, or farther from, the desired endpoint. This helps the agent to figure out 
where it should go.
There is an inherent asymmetry between the two constraints the agent has to consider as it finds its path
towards the endpoint. While a collision of the path with the branch cut can be detected, and punished, immediately,
the information of whether the contour is continuously deformable into the original contour in the absence
of the branch cut can only be provided once the agent has reached its goal. It thus receives the respective
reward or punishment only at the very end of an episode, which makes it much harder for the agent to figure
out which of the steps that it took was responsible for the reward or punishment it received in the end.
This is the aforementioned problem of sparse rewards and poses a challenge that has to be addressed.  


\section{\label{sec:experiments}Numerical Experiments and Results}
In Section \ref{sec:rew} above we discussed the issue of the pole detection being particularly difficult to
learn for an agent, as the reward or punishment is only provided at the end of the episode. We thus decided
to study two separate cases. First, we investigated scenarios that featured branch cuts only, that is, we 
neglected the presence of the poles. This scenario is easier for an agent, as it is provided with rewards and
punishments immediately, that is, after every single step. Its actions thus have very immediate consequences, 
and it is easy for the agent to derive optimal actions. In a second scenario, we consider the full problem, 
that is, besides the branch cut we also introduce the poles. This has a severe effect on the agents performance,
as the reward signal is now delayed. As a possible simplification of the active pole detection scenario we 
also introduced a mechanism that allows the environment to automatically close the contour when the agent has 
successfully exited the branch cut structure. We refer to this mechanism as path auto-completion.
The training of the agents has been conducted as follows. For each episode, we pick a value for $x$ at random,
by uniformly sampling from the plane $x\in [-5,5]$ GeV$^2 \times i[-5,5]$ GeV$^2$. The position of the poles,
if present, is fixed at $\pm\frac{i}{2}$ GeV$^2$.

The policy network (actor) and value estimator network (critic) were chosen identically. They consist of two hidden layers, 
the first with $64$ neurons and the second with $32$ units. The activation function for each 
layer is the hyperbolic tangent, and as optimizer we used Adam \cite{Kingma:2014aa}. 
An overview of the other parameters is summarized in Table \ref{tab:experiment_parameters}.

All numerical experiments were conducted on Nvidia Titan RTX Graphics Processing Units (GPUs). On three
GPUs, we could conduct $48$ experiments in parallel. The program code has been written in Python, using PyTorch 
\cite{Paszke:2019aa} with GPU support.

\begin{widetext}
\begin{center}
\begin{table}
    \begin{tabular}{|c|c|c|}        
        \hline
        \textbf{property}       & \textbf{values}        &   \textbf{description} \\
        \hline
        State space dim.    & $14$  &  number of parameters in state space\\
        Action space dim.    & $2$  &  dimension for step length and angle\\
        \hline
        First pole    & $0 + 0.5i$  &  position first pole\\
        Second pole    & $0 - 0.5i$  &  position second pole\\
         \hline
        $ |x| $        & $\mathbb{R}$  &  modulus of ext. momentum square \\
        \hline        
        $ \arg x $     & $[-\pi,\pi]$    &  argument of ext. momentum square \\
        \hline
        $ \Lambda^2 $  & $10$  &  cut-off on real axis (contour endpoint)\\
        \hline
        $\epsilon$-region  & $0.5$  &  radius around goal\\
        \hline
        $ \gamma$  & $0.99$  &  discount factor\\
        \hline
        eps\_clip  & $0.2$  &  clip parameter for PPO\\
        \hline
        batch\_size  & $80$ &  batch size\\
        \hline
        update\_steps  & $6000$ &  time steps between policy updates\\
        \hline
        learning\_rate  & $5\times 10^{-5}$ &  learning rate for optimizer\\
        \hline
        act\_std  & $0.4$  &  standard deviation for action distribution (multivariate normal)\\
        \hline
        rew\_fail  & $-50000$  &  reward for failure\\
        rew\_coll  & $-10$  &  reward for collision with contour\\
        rew\_iter  & $-5$  &  reward for each step\\
        rew\_attr  & $\{.01,\ .0055,\ .001\}$  &  reward for goal attraction\\
        rew\_won  & $1000$  &  reward for winning\\
        \hline
        max episodes  & $5000$  &  max number of episodes for a single policy training\\
        max steps    & $1500$  &  max number of steps per episode\\
        \hline
    \end{tabular}
    \caption{Parameters of the three numerical experiments without pole detection as described in Section \ref{sec:reference_experiments}. 
    The same parameter sets have also been used in the pole detection runs with auto-completion, see Section \ref{sec:pole_detection_w_auto_completion}. 
    The only difference between the three runs is in the parameter \texttt{rew\_attr}, 
    which controls the reward or punishment received if the agent steps closer to, or away from, the desired endpoint.}
    \label{tab:experiment_parameters}
\end{table}
\end{center}
\end{widetext}

\subsection{\label{sec:reference_experiments}Experiments without pole detection}
We conducted our first numerical experiments with the simplified environment, that is, we ignored the presence of the poles.  
We considered three configurations that differ in the choice for the parameter \texttt{rew\_attr}. 
This parameter is responsible for controlling the reward (or punishment) the agent receives as it
steps closer to, or farther from, the desired endpoint. The three parameter values were
$\{.01,\ .0055,\ .001\}$. All other parameters were identical to the values shown in 
Table \ref{tab:experiment_parameters}.

\begin{figure*}[t]
\centering
\subfigure[][]{
 \label{fig:ref_exp_a}
\includegraphics[width=0.45\hsize]{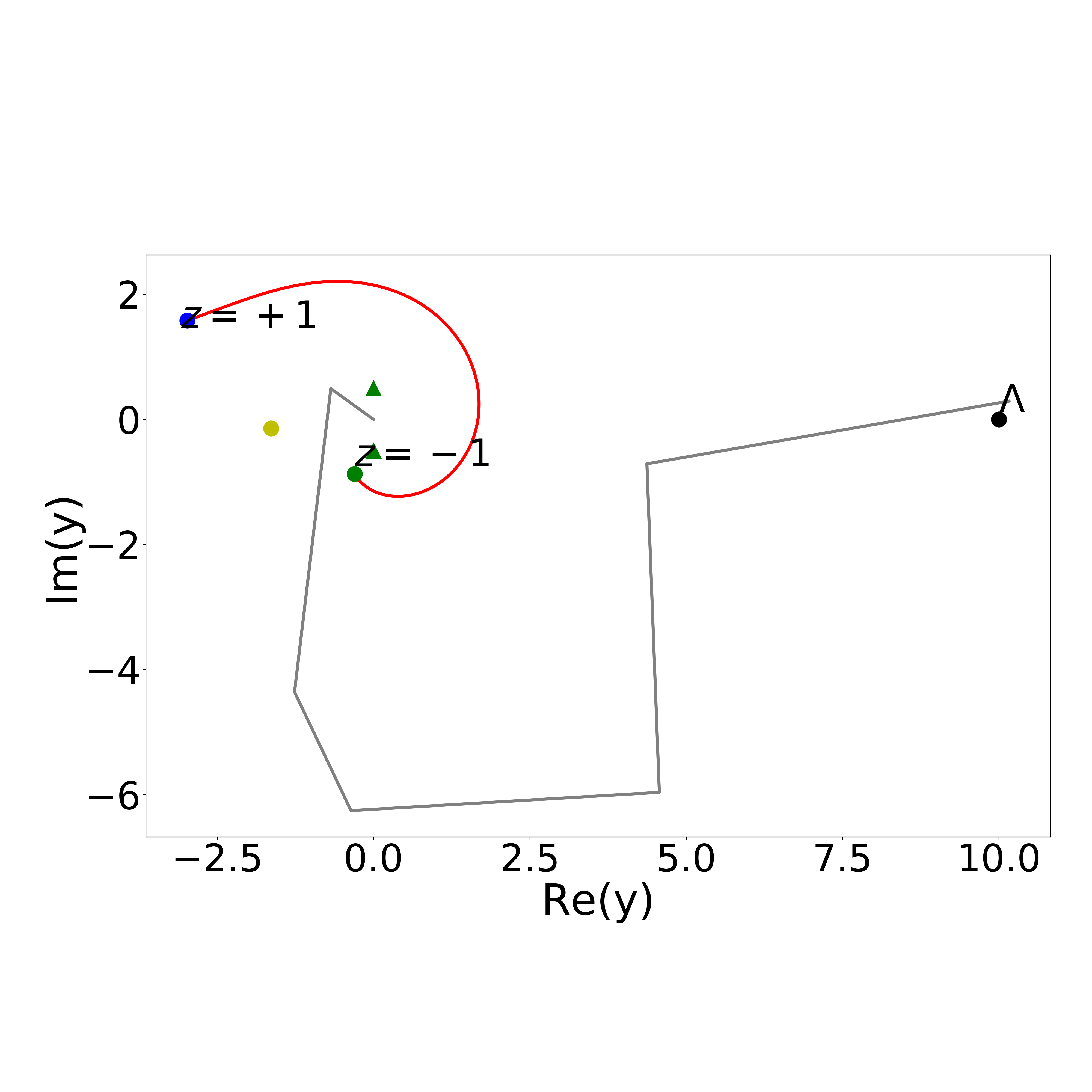}
}\hspace{8pt}
\subfigure[][]{
 \label{fig:ref_exp_b}
\includegraphics[width=0.45\hsize]{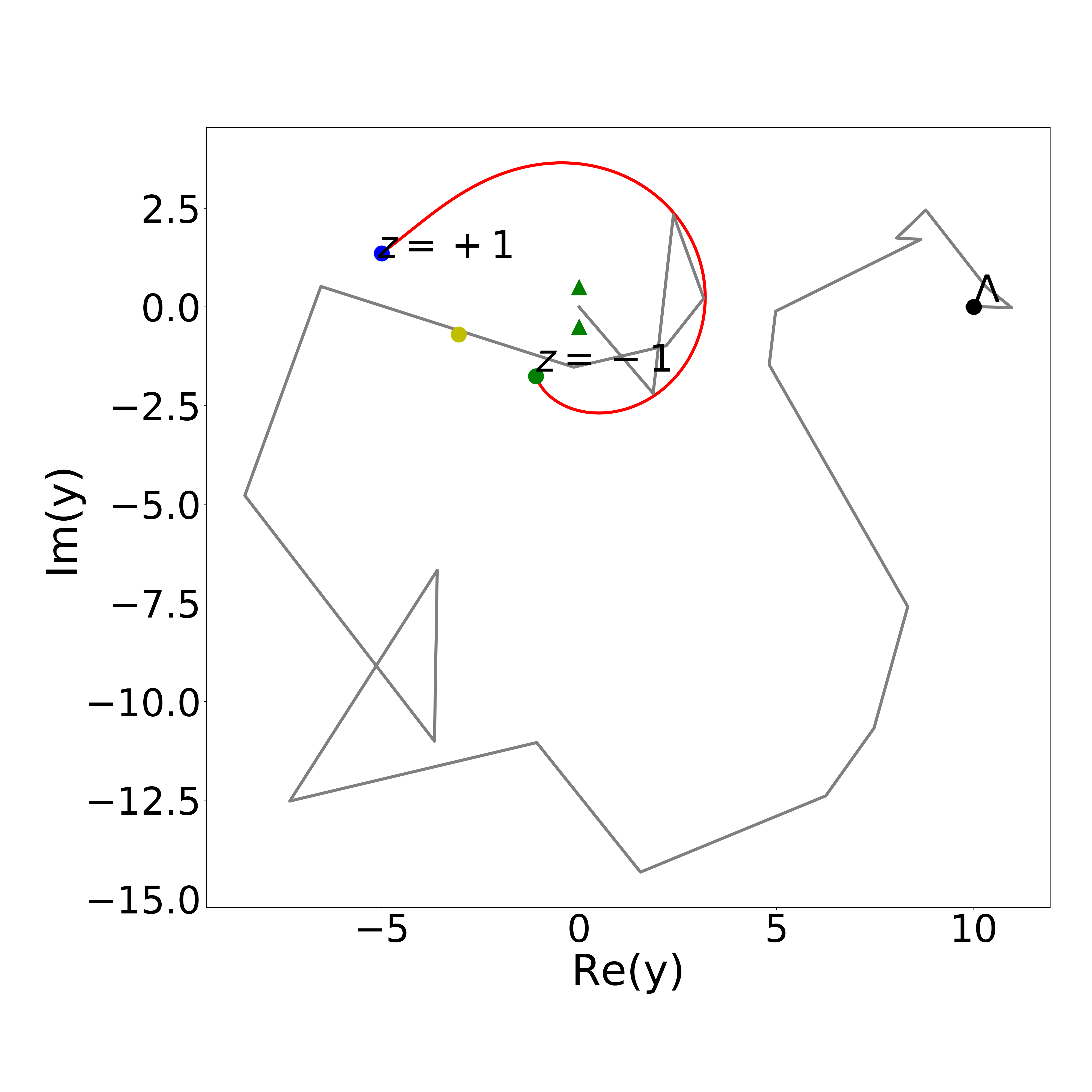}
}
\caption[]{Two solutions provided by the same agent throughout training. Note that we do not apply any smoothing mechanisms, 
nor do we discourage or remove loops in the paths produced by the agent. Each contour starts at the origin and traces 
out a path that leads to the cutoff labeled $\Lambda$. The (green) triangles mark the position of the poles, which were ignored 
in this particular run. The (yellow) circles between the cut's endpoints indicate the value of $x$ used for this particular run. The solutions presented here 
were produced by the best performing agent among the three trained in this setting. It has a success rate of almost $96\%$. 
Note that both contours solve the problem. 
Figure \subref{fig:ref_exp_a} shows a valid solution produced by the agent after training for 1783 episodes. 
Figure \subref{fig:ref_exp_b} shows a valid solution produced by the agent after training for 4726 episodes. This path could be post-processed
to remove the loops, which we haven't implemented.}
\label{fig:ref_exp}
\end{figure*}

\begin{figure}
    \centering 
    \includegraphics[width=\columnwidth]{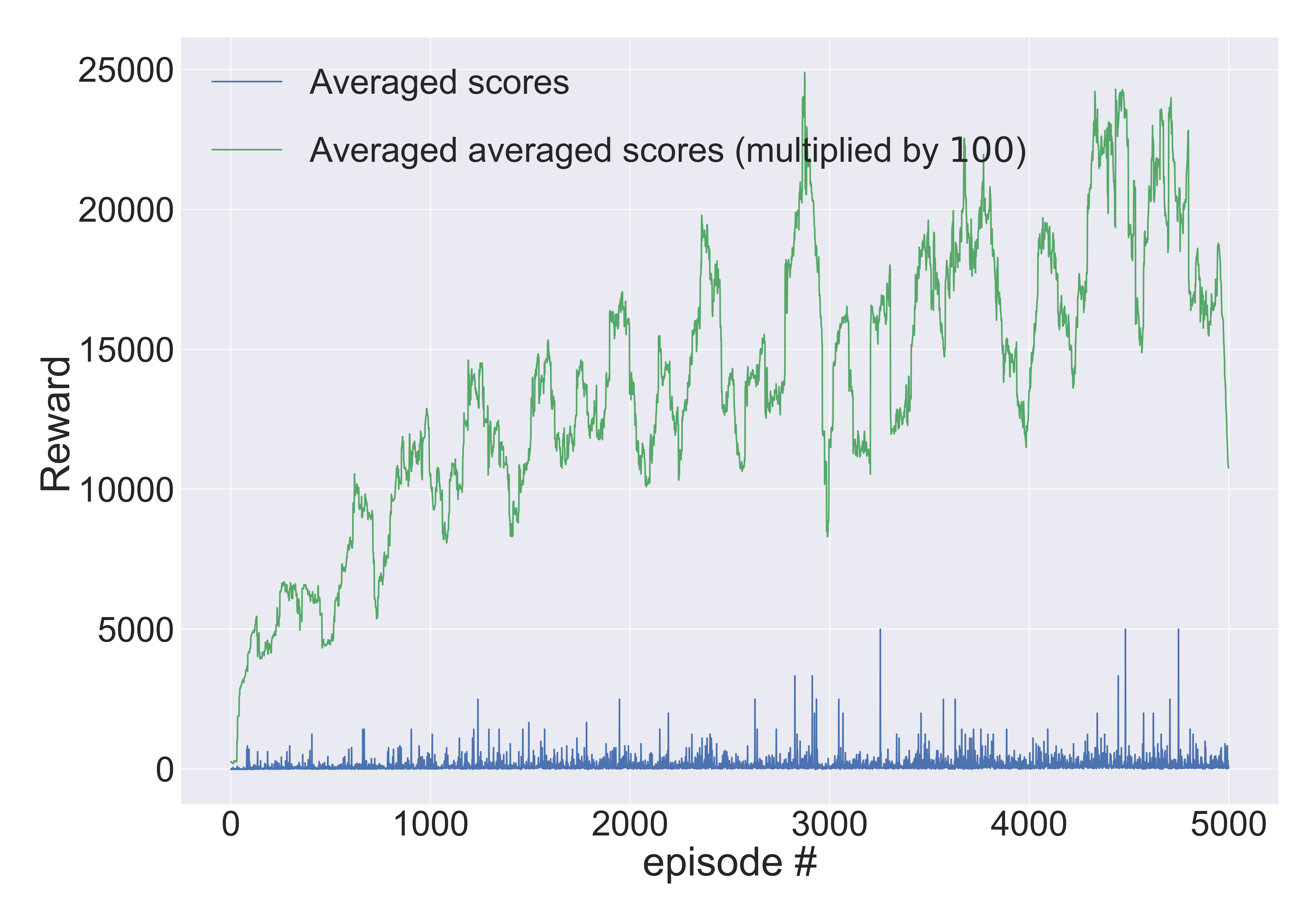}
    \caption{This plot illustrates the development of averaged rewards throughout training of the best reference experiment, where the pole detection was deactivated. The blue curve represents the averaged return for each episode, while the green curve is the average over $100$ consecutive averaged returns, rescaled for readability. The trend of improvement of the performance over training time is clearly visible.}
    \label{fig:reference_results_learning_curve}
\end{figure}

The outcome of all three experiments lie above a 90 percent success rate. For the experiment 
with an attractor reward of $0.01$ we get a success rate of $92.94\%$, with an attractor reward of 
$0.0055$ we get $94.56\%$, and with an attractor reward of $0.001$ we get a success rate of $95.58\%$. 
Figure~\ref{fig:ref_exp} illustrates two non-trivial scenarios where the agent 
(in this case the agent with a final success rate of $95.58\%$) was able to reach a point within
a disc with radius $\epsilon$\texttt{-region}, centered around the end-point. 
Figure~\ref{fig:reference_results_learning_curve} shows the learning curve of the best agent trained in 
this environment.

\subfiglabelskip=0pt

\subsection{\label{sec:pole_detection}Experiments with pole detection}
\begin{figure*}[t]
\centering
\subfigure[][]{
 \label{fig:pole_exp_a}
\includegraphics[width=0.45\hsize]{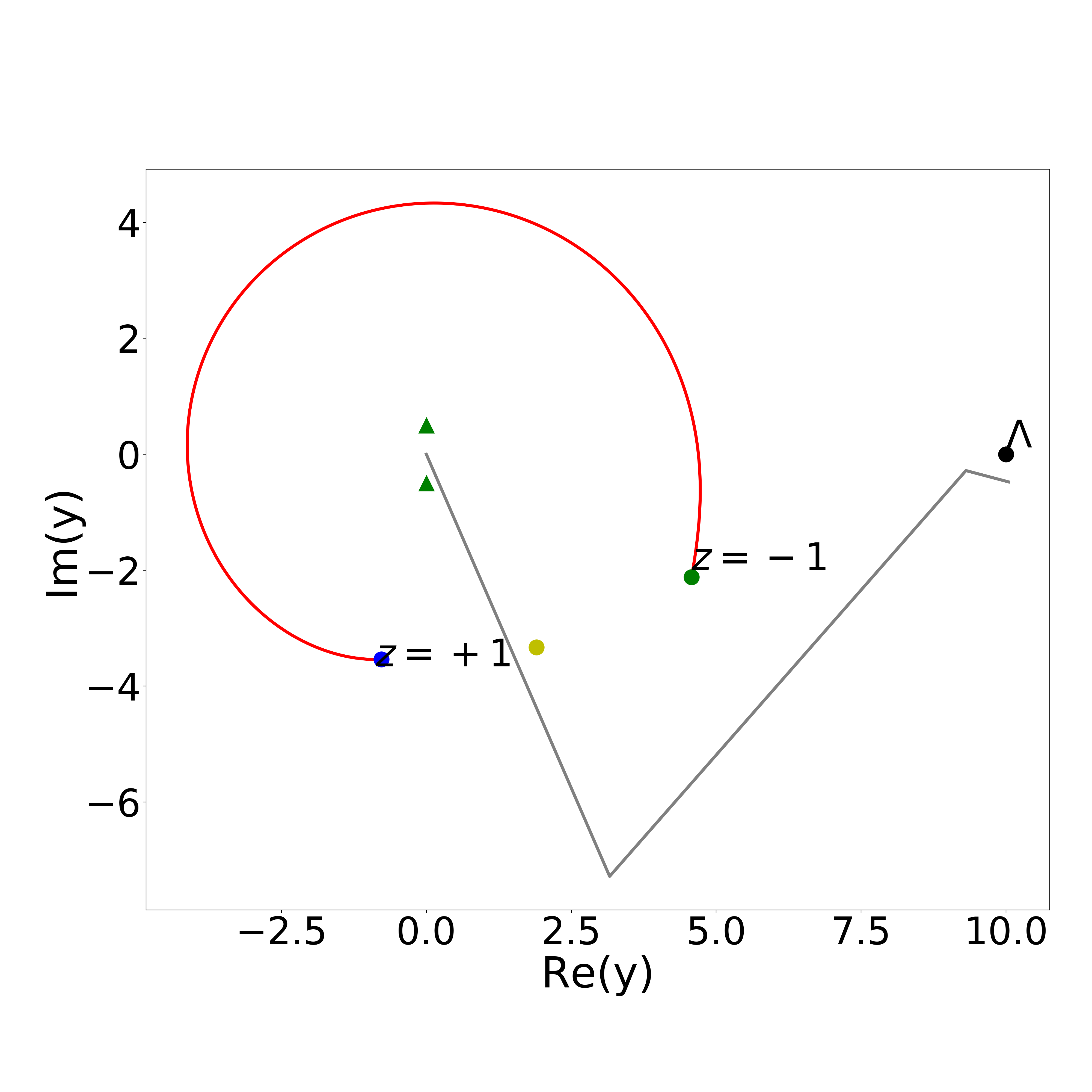}
}\hspace{8pt}
\subfigure[][]{
 \label{fig:pole_exp_b}
\includegraphics[width=0.45\hsize]{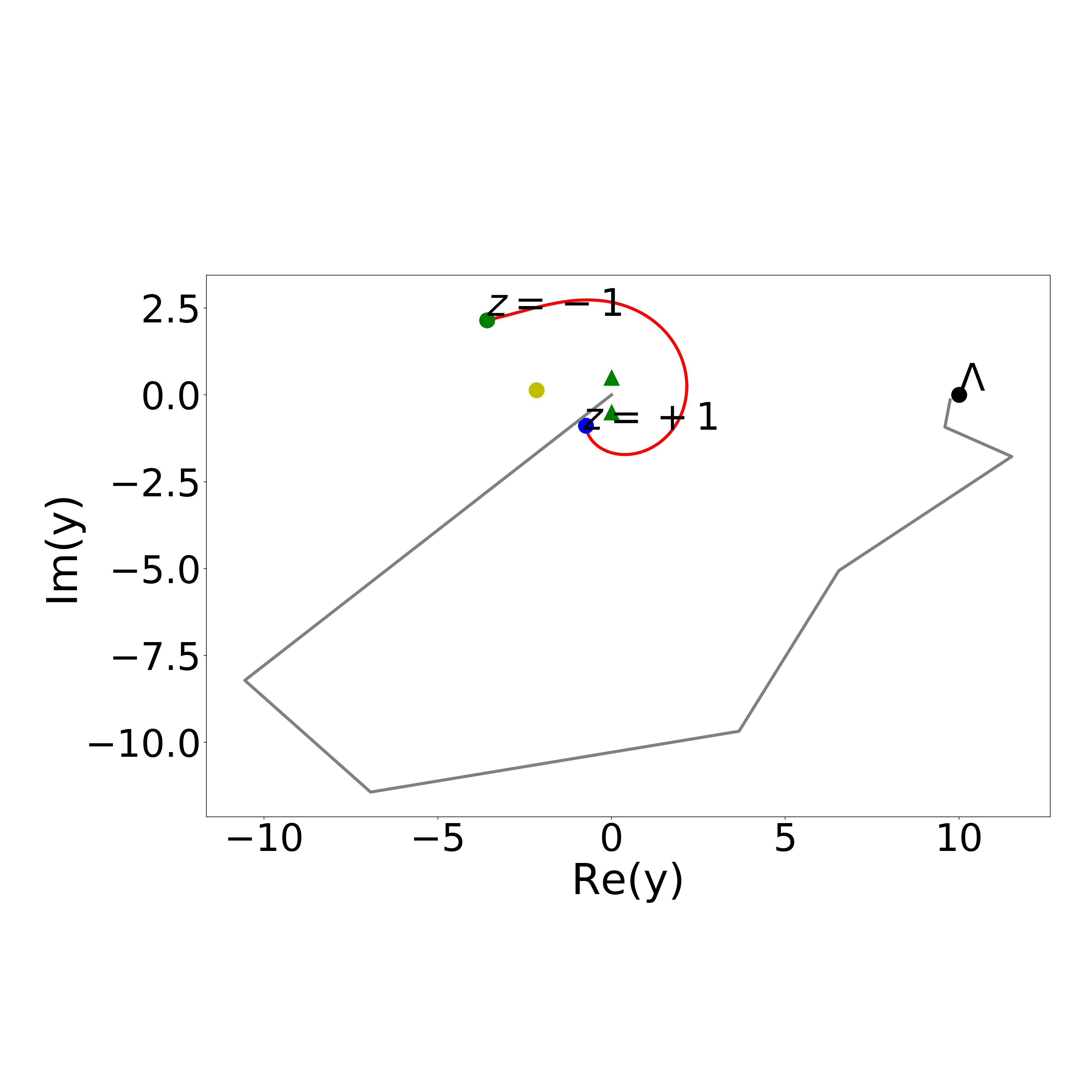}
}
\caption[]{Two solutions suggested by the same agent throughout training on the full problem (poles included). The (green) triangles mark the position of the poles. 
The solutions presented here were produced at a late time throughout training by the best performing agent among the 720 trained in this setting. 
It has a success rate of almost $28\%$, which is considerably lower than the one achieved without pole detection.
Figure \subref{fig:pole_exp_a} shows a valid solution produced by the agent after training for 4996 episodes.
Figure \subref{fig:pole_exp_b} shows an invalid solution produced by the agent after training for 4674 episodes. This contour must be rejected, because
the path is not continuously deformable into the path along the positive real half-axis in absence of the branch cut.}
\label{fig:pole_exp}
\end{figure*}

\begin{figure}
    \centering 
    \includegraphics[width=\columnwidth]{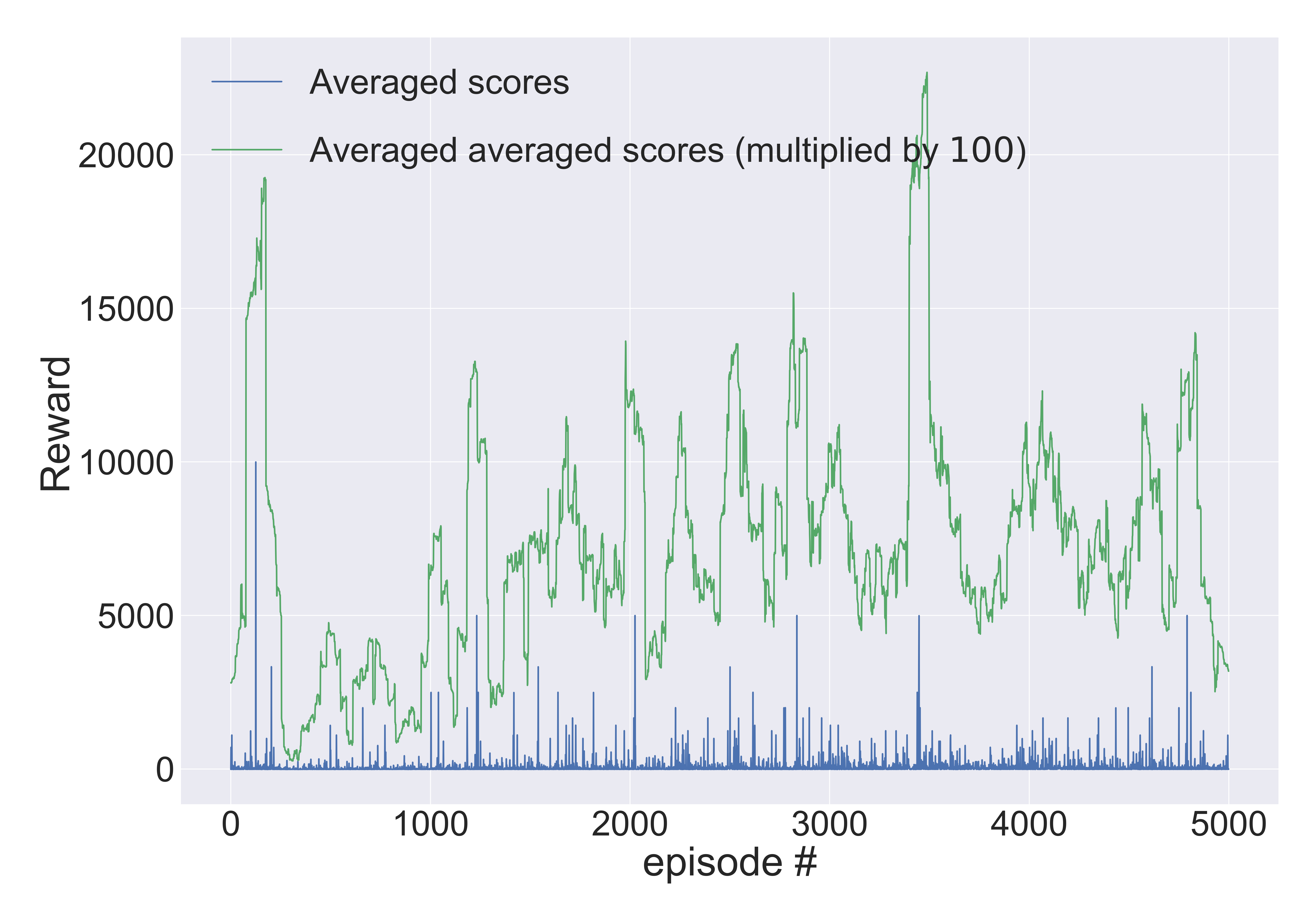}
    \caption{This plot illustrates the development of averaged rewards throughout training of the best experiment with pole detection. The blue curve represents the averaged return for each episode, while the green curve is the average over $100$ consecutive averaged returns, rescaled for readability.}
    \label{fig:pole_detection_results_learning_curve}
\end{figure}

To investigate the performance of a reinforcement learning agent on the full problem,
that is, with poles included, we conducted a total of $720$ numerical experiments with different parameters. 
The parameters that we varied were the following. We chose different values for the standard deviation
of the noise distribution used for the actions (\texttt{act\_std}), for the number of steps in between
the updates of the policy (\texttt{update\_steps}), for the learning rate used by the optimizer 
(\texttt{learning\_rate}), the batch size (\texttt{batch\_size}), and for the parameter 
\texttt{rew\_attr} that was also varied throughout the experiments without the poles being present.
The overall 720 numerical experiments are all combinations of:
\begin{itemize}
\item \texttt{act\_std} $\in\{.1,\ .2,\ .3,\ .4\}$ 
\item \texttt{update\_steps} $\in\{250,\ 562,\ 875,\ 1187,\ 1500\}$ 
\item \texttt{learning\_rate} $\in\{10^{-5},\ 10^{-4},\ 10^{-3}\}$ 
\item \texttt{batch\_size} $\in\{60,70,80,90\}$ 
\item \texttt{rew\_attr} $\in\{.01,\ .0055,\ .001\}$ 
\end{itemize}
The best experiment with activated pole detection achieved a success rate of $27.82\%$.
The parameters associated with the best performing agent were: 
\begin{itemize}
\item \texttt{act\_std}: $.4$ 
\item \texttt{update\_steps}: $1187$ 
\item \texttt{learning\_rate}: $10^{-5}$ 
\item \texttt{batch\_size}: $90$ 
\item \texttt{rew\_attr}: $.0055$ 
\end{itemize}
This is significantly smaller than the success rates achieved by the agents trained in the environment without pole detection. 
Figure~\ref{fig:pole_exp} shows two situations, one with positive and one with negative outcome. In contrast to the 
simplified experiments discussed in the previous section, the agent cannot always count reaching the goal within a given tolerance as success in this case. 
Positive experiences are much less frequent, which slows down the learning process. The learning curve of the best agent trained in this scenario 
is shown in Figure \ref{fig:pole_detection_results_learning_curve}

\subfiglabelskip=0pt

\subsection{\label{sec:pole_detection_w_auto_completion}Experiments with pole detection and auto-completion}
\begin{figure*}[t]
\centering
\subfigure[][]{
 \label{fig:pole_exp_autoc_a}
\includegraphics[width=0.42\hsize]{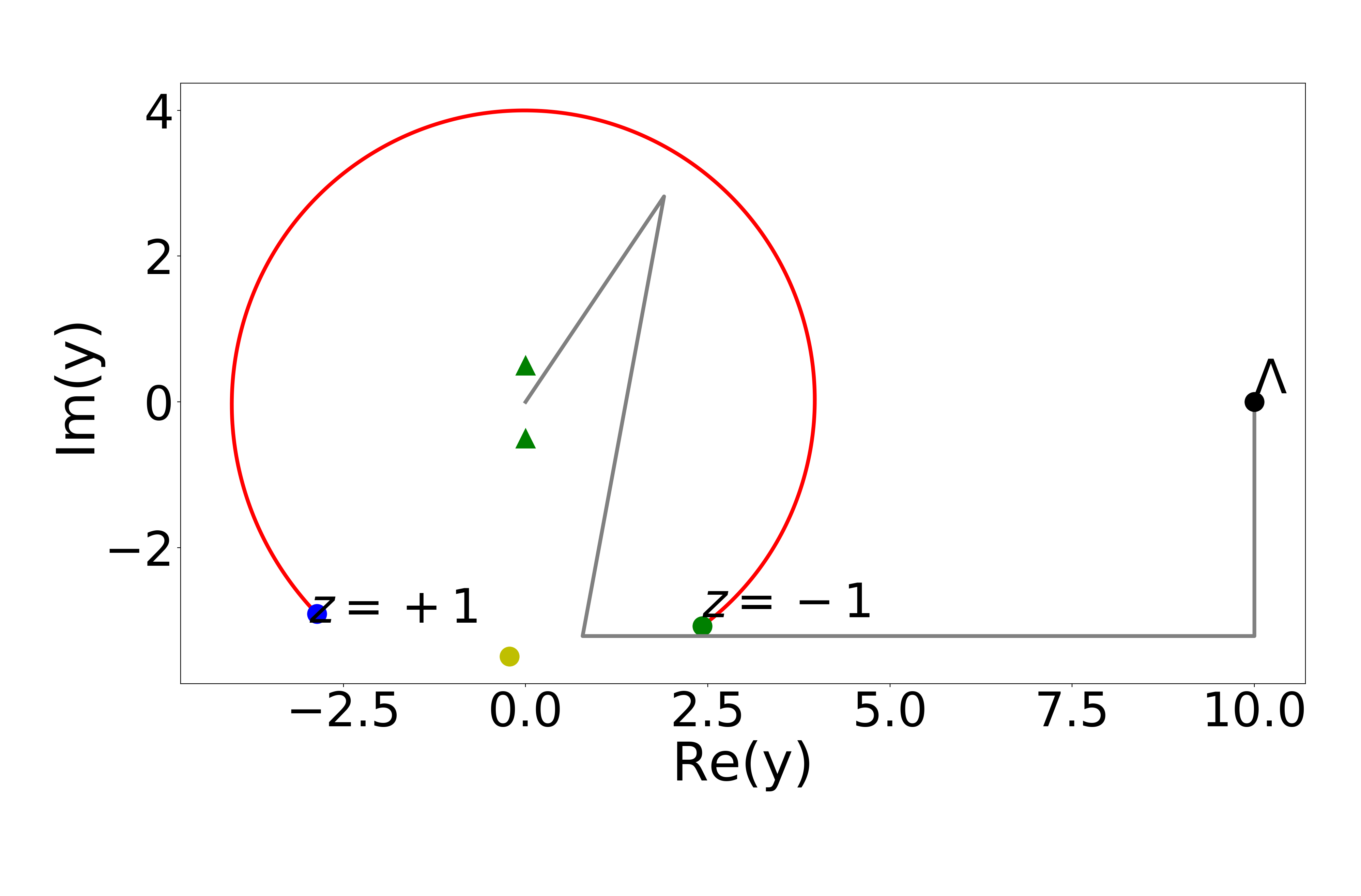}
}\hspace{8pt}
\subfigure[][]{
 \label{fig:pole_exp_autoc_b}
\includegraphics[width=0.42\hsize]{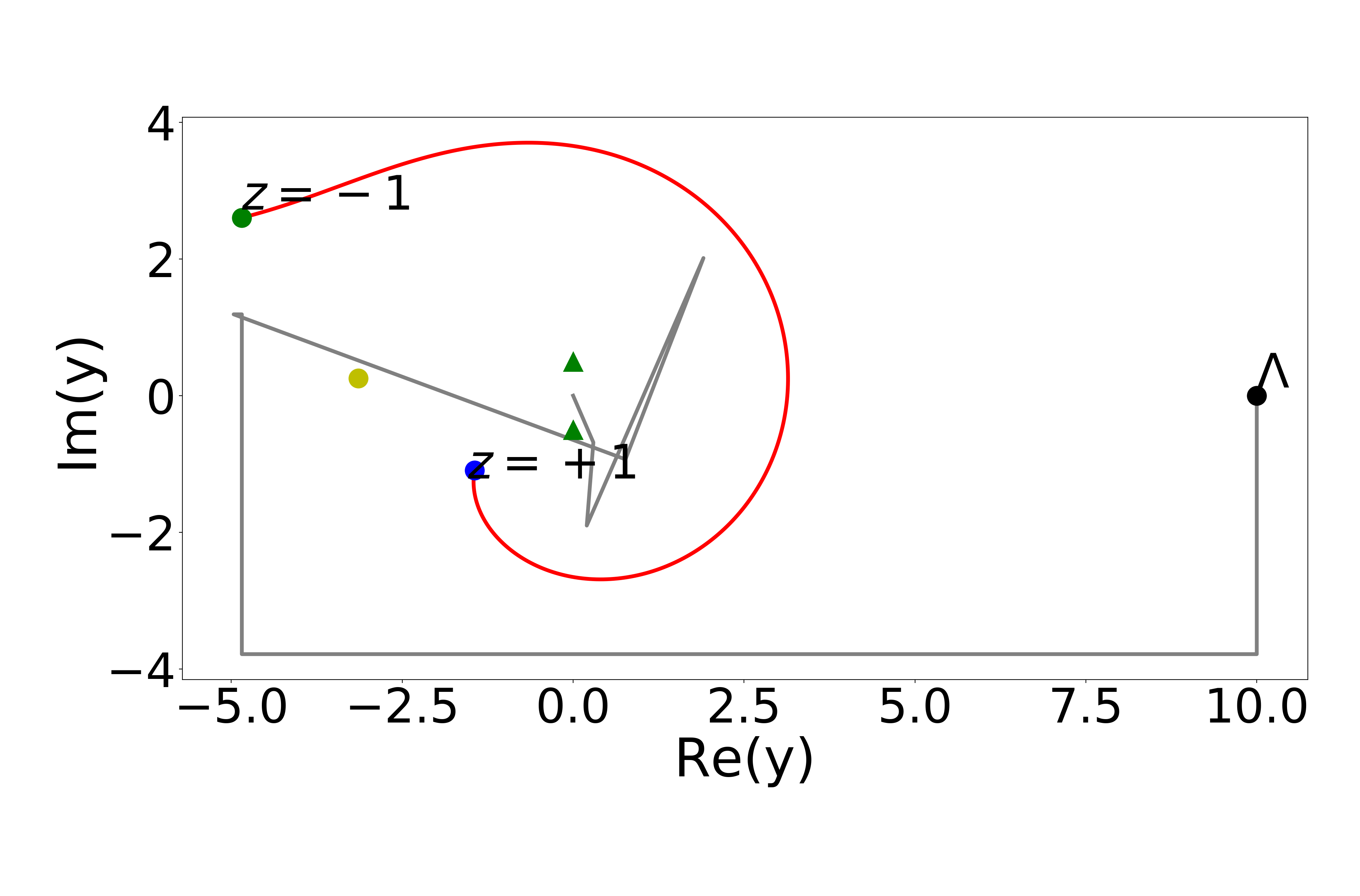}
}
\caption[]{Two solutions suggested by the same agent throughout training on the full problem (poles included) with auto-completion. 
The (green) triangles mark the position of the poles. The solutions presented here were produced by the best agent throughout training. 
It has a success rate of almost $48\%$, which is a dramatic increase in performance as compared to the best agent trained with naive pole detection. 
This indicates that this issue is due to a delayed reward signal.
Figure \subref{fig:pole_exp_autoc_a} shows a valid solution produced by the agent after training for 126 episodes.
Figure \subref{fig:pole_exp_autoc_b} shows a valid solution produced by the agent after training for 1820 episodes.}
\label{fig:pole_exp_autoc}
\end{figure*}

In order to make the reward, or punishment, for enclosing a pole more immediate to the agent, 
we decided to alter the environment as follows. Rather than allowing the agent to move freely 
within the plane  $x\in [-1.5\Lambda^2,1.5\Lambda^2]$ GeV$^2 \times i[-1.5\Lambda^2,1.5\Lambda^2]$ GeV$^2$, 
we first drew a bounding box around the branch cut. The size of the box was chosen such that it confines the cut as close as
possible, without actually intersecting it. One can then make the box slightly larger such that 
its edges can be used as part of the integration contour. This slight modification is necessary, as
the numerical integration becomes increasingly expensive with decreasing distance to the branch cut.
The agent was then trained to produce a contour that 
connects the origin with the box that encloses the branch cut. Once such a contour exists, where the opening 
of the branch cut has to be respected, that is, no intersections with the cut are allowed, the auto-completion 
algorithm takes over. The contour is then closed automatically by following along the edges of the box and then parallel to the real axis, 
until the real part of the contour is equal to $\Lambda^2$, after which the contour is closed by a vertical line segment.  
Note that the auto-completion algorithm also checks for the location of the poles.   
Since the potential pickup of a pole's residue depends upon whether 
the auto-completed contour is closed by following the edges to the left, or by following the edges 
to the right of the intersection point, or cannot be avoided regardless of how the contour is closed, 
the algorithm proceeds as follows. If the agent reaches the edge of the bounding box in the second or third 
quadrant, we test both ways of closing the contour and accept the one that doesn't pick up the residue, if it exists. 
This setup improves upon the delayed reward problem, as the auto-completion is -- apart from the direction 
of the closure -- beyond the agent's control, 
whose responsibility ends with reaching one of the edges of the bounding box. This makes the impact of its 
choices much more immediate. Consequently, we found a much better performance of the agent working on 
the full problem. We conducted three experiments, using the same parameters as the ones used for the 
simplified environment shown in Table \ref{tab:experiment_parameters}. 
The best success rate was around $47\%$, produced by the agent using a \texttt{rew\_attr} of $0.001$. 
Note furthermore, that this approach can also be adopted for the Dyson-Schwinger equation setting. 
Figure~\ref{fig:pole_exp_autoc} shows two contours suggested throughout training by an agent  learning in this environment.

\subfiglabelskip=0pt

\section{\label{sec:summary}Summary and outlook}
In this paper we introduced a novel technique based on deep reinforcement learning that allows 
for numerical analytic continuation of loop integrals in Euclidean space in quantum field theory.
The main challenge to solve such integrals in the complex domain is to find suitable contours that
avoid branch cuts and poles in the integration plane. By referring to an example where contour deformation
has been successfully applied to solve the problem in the past, we trained reinforcement learning 
agents to produce valid contours for the same example, providing a proof of principle that machine learning can indeed
be of assistance in solving such tedious problems.  

While the best agent trained in the absence of poles reaches success rates that justify deployment in such scenarios, 
agents trained on settings with poles being present achieved success rates that produce valid contours only for every other 
value of $x$. Since we did not directly address the issue of delayed rewards in this first study, we are 
confident that this success rate can be increased sufficiently to allow for deployment of this approach in 
the presence of poles as well.

In a follow-up study we will thus focus on two main tasks. 
First, we will improve upon the agent's performance in the presence of poles, which is
a scenario in which we have to deal with sparse rewards. For studies addressing this issue see e.~g.~
\cite{Jaderberg:2016aa,Andrychowicz:2017aa,Pathak:2017aa}. We will explore the 
applicability of the strategies presented in these studies to our problem (note that, e.~g.~\cite{Andrychowicz:2017aa}
requires an off-policy approach). We will also investigate how an ensemble of trained agents performs at solving the problem
of producing valid contours in the presence of poles. Since we can validate each contour by means of the auxiliary
function used throughout training, we could then just ask every agent in the ensemble to produce a contour and 
accept one that is valid. If, on the other hand, the ensemble is incapable of producing a valid contour for a 
given set of parameters, this will give us insights as to why the agents fail to solve that particular 
situation, such that we can use domain knowledge to provide guidance. In addition to looking into approaches that try 
to improve upon the sparse reward problem, as well as investigating ensembles of agents, we furthermore will 
also explore the potential of runtime enforcement (shielding) \cite{Alshiekh:2018aa} for our application. 
 
Another important question is, how the environment can be generalized in order to being
able to deal with integral equations of the Fredholm 2 type, such as Dyson-Schwinger equations. 
In order to address this, we plan to take numerical input of the complex integration plane before an iteration step is conducted. 
Here are two possibilities as of how the perception of the
non-analyticities is approached. One could use conventional image detection algorithms in combination with the theory
of complex functions to detect poles and branch cuts (see e.~g.~\cite{Windisch:2016iud} for such an approach to detect
poles) in the integration plane. Instead of conventional image processing one could also work with convolutional
neural networks on a `pixel' basis, where the pixels are in fact the moduli of the numerically computed
complex values of the integrand. 

Once the information about the obstructions in the integration plane is available, it can be fed into the 
vectorized environment presented here. We can then sample the integration plane for various values of $x$ and 
train the agent in a similar fashion. Thus, every iteration step corresponds to the 
exact same problem as the one treated in this paper. Consequently, the agent has to be re-trained for every iteration 
step. One can, however, start with the policy the agent used for the previous iteration step, and if the shift in 
the analytic properties in the integration plane is not too severe, the agent should be able to adapt easily. 
Another aspect that should help in this regard is, that we formulated the environment in a very general way.
For example, it is not expected that branch cuts arising in such integrals open in directions other than $\arg{x}$,
but they will most likely just differ by its shape. The difference in shape will force the agent to focus more
on the direction of the opening until it leaves the structure. 

But even assuming that both strategies, the one directly addressing the sparse reward problem and the ensemble approach, 
fail to produce a valid answer for a certain point of $x$ in a Dyson-Schwinger equation setting, the consequences 
are not expected to be severe, since the system is solved iteratively, so even if an iteration step fails, the system
could still converge, given that the overall success rate is not too low.

With a success rate of $96 \%$ without poles, and close to $50 \%$ with poles being present (and without applying any
of the delayed reward signal strategies mentioned above), our study shows that 
machine learning in general, and reinforcement learning in particular, can help in tackling such demanding numerical 
problems as contour deformations to compute one-loop integrals in quantum field theory.

\section{Acknowledgments}
This work has been supported by the Silicon Austria Labs (SAL) funded by the Austrian
federal government and the federal states of  Carinthia, Styria and Upper Austria and
Austrian Association for the Electric and Electronics Industry.
Furthermore, AW acknowledges that parts of this work have been supported by the U.S. Department of Energy, 
Office of Science, Office of Nuclear Physics under Award Number \#DE-FG-02-05ER41375, 
as well as by the Austrian Science Fund (FWF), Schr\"odinger Fellowship J3800-N27.  
\appendix
\section{\label{app:A}Conventions}
All calculations are performed in Euclidean space.
\subsection{Hyperspherical coordinates}

Using hyperspherical coordinates, the 4-momentum $q$ can be expressed as

\begin{eqnarray}
&&\int_{\mathbb{R}^{4}}d^{4}q \rightarrow\\ 
&&\int_{0}^{2\pi}d\phi\int_{0}^{\infty}dq\ q^{3}\int_{0}^{\pi}d\theta_{1}\sin^{2}\theta_{1}\int_{0}^{\pi}d\theta_{2}\sin\theta_{2}\label{eq:Convention7}\nonumber\\
 & = & \Bigg|\begin{array}{c}
y\equiv q^{2}\rightarrow dy=2qdq\nonumber\\
\theta_{1}\equiv\arccos z\rightarrow d\theta_{1}=-\frac{dz}{\sqrt{1-z^{2}}}\\
\theta_{2}\equiv\arccos w\rightarrow d\theta_{2}=-\frac{dw}{\sqrt{1-w^{2}}}
\end{array}\Bigg|\\
 & = & \frac{1}{2}\int_{0}^{2\pi}d\phi\int_{0}^{\infty}dy\ y\int_{-1}^{1}dz\sqrt{1-z^{2}}\int_{-1}^{1}dw.\nonumber 
\end{eqnarray}

With two momenta involved, only the radial and one angular integral remains. With an IR cutoff $\varepsilon$ and an UV cutoff $\Lambda$, we get

\begin{eqnarray}
&&\int_{\mathbb{R}^{4}}\frac{d^{4}q}{(2\pi)^4} \rightarrow\\ 
&&\frac{1}{(2\pi)^3}\int_{\varepsilon}^{\Lambda}dy\ y\int_{-1}^{1}dz\sqrt{1-z^{2}}\nonumber.
\end{eqnarray}

Defining $x$ to be the square of the external momentum $p^2$,

\begin{eqnarray}
x := p^2,
\end{eqnarray}

the scalar product between $p$ and $q$ becomes.

\begin{eqnarray}
p.q = \sqrt{x}\sqrt{y}z.
\end{eqnarray}


\bibliographystyle{utphys_mod}
\bibliography{citations}

\providecommand{\href}[2]{#2}\begingroup\raggedright\begin{thebibliography}{10}

\bibitem{Alkofer:2000wg}
R.~Alkofer and L.~von Smekal,
  \href{http://dx.doi.org/10.1016/S0370-1573(01)00010-2}{{\em Phys. Rept.} {\bf
  353} (2001)  281},
\href{http://arxiv.org/abs/hep-ph/0007355}{{\tt arXiv:hep-ph/0007355
  [hep-ph]}}.

\bibitem{Fischer:2006ub}
C.~S. Fischer, \href{http://dx.doi.org/10.1088/0954-3899/32/8/R02}{{\em J.
  Phys.} {\bf G32} (2006)  R253--R291},
\href{http://arxiv.org/abs/hep-ph/0605173}{{\tt arXiv:hep-ph/0605173
  [hep-ph]}}.

\bibitem{Roberts:2007jh}
C.~D. Roberts, M.~S. Bhagwat, A.~Holl, and S.~V. Wright,
  \href{http://dx.doi.org/10.1140/epjst/e2007-00003-5}{{\em Eur. Phys. J. ST}
  {\bf 140} (2007)  53--116},
\href{http://arxiv.org/abs/0802.0217}{{\tt arXiv:0802.0217 [nucl-th]}}.

\bibitem{Sanchis-Alepuz:2015tha}
H.~Sanchis-Alepuz and R.~Williams,
  \href{http://dx.doi.org/10.1088/1742-6596/631/1/012064}{{\em J. Phys. Conf.
  Ser.} {\bf 631} (2015) no.~1, 012064},
\href{http://arxiv.org/abs/1503.05896}{{\tt arXiv:1503.05896 [hep-ph]}}.

\bibitem{Cloet:2013jya}
I.~C. Cloet and C.~D. Roberts,
  \href{http://dx.doi.org/10.1016/j.ppnp.2014.02.001}{{\em Prog. Part. Nucl.
  Phys.} {\bf 77} (2014)  1--69},
\href{http://arxiv.org/abs/1310.2651}{{\tt arXiv:1310.2651 [nucl-th]}}.

\bibitem{Eichmann:2016yit}
G.~Eichmann, H.~Sanchis-Alepuz, R.~Williams, R.~Alkofer, and C.~S. Fischer,
  \href{http://dx.doi.org/10.1016/j.ppnp.2016.07.001}{{\em Prog. Part. Nucl.
  Phys.} {\bf 91} (2016)  1--100},
\href{http://arxiv.org/abs/1606.09602}{{\tt arXiv:1606.09602 [hep-ph]}}.

\bibitem{Sanchis-Alepuz:2017jjd}
H.~Sanchis-Alepuz and R.~Williams,
  \href{http://dx.doi.org/10.1016/j.cpc.2018.05.020}{{\em Comput. Phys.
  Commun.} {\bf 232} (2018)  1--21},
\href{http://arxiv.org/abs/1710.04903}{{\tt arXiv:1710.04903 [hep-ph]}}.

\bibitem{Eichmann:2019vhk}
G.~Eichmann, ``{Towards resonance properties in the Dyson-Schwinger
  approach},''
\newblock 2019.
\newblock
\href{http://arxiv.org/abs/1912.08873}{{\tt arXiv:1912.08873 [hep-ph]}}.
\newblock

\bibitem{Windisch:2013dxa}
A.~Windisch, M.~Q. Huber, and R.~Alkofer,
  \href{http://dx.doi.org/10.5506/APhysPolBSupp.6.887}{{\em Acta Phys. Polon.
  Supp.} {\bf 6} (2013) no.~3, 887--892},
\href{http://arxiv.org/abs/1304.3642}{{\tt arXiv:1304.3642 [hep-ph]}}.

\bibitem{Maris:1995ns}
P.~Maris, \href{http://dx.doi.org/10.1103/PhysRevD.52.6087}{{\em Phys. Rev.}
  {\bf D52} (1995)  6087--6097},
\href{http://arxiv.org/abs/hep-ph/9508323}{{\tt arXiv:hep-ph/9508323
  [hep-ph]}}.

\bibitem{Strauss:2012dg}
S.~Strauss, C.~S. Fischer, and C.~Kellermann,
  \href{http://dx.doi.org/10.1103/PhysRevLett.109.252001}{{\em Phys. Rev.
  Lett.} {\bf 109} (2012)  252001},
\href{http://arxiv.org/abs/1208.6239}{{\tt arXiv:1208.6239 [hep-ph]}}.

\bibitem{Windisch:2012zd}
A.~Windisch, R.~Alkofer, G.~Haase, and M.~Liebmann,
  \href{http://dx.doi.org/10.1016/j.cpc.2012.09.003}{{\em Comput. Phys.
  Commun.} {\bf 184} (2013)  109--116},
\href{http://arxiv.org/abs/1205.0752}{{\tt arXiv:1205.0752 [hep-ph]}}.

\bibitem{Windisch:2012sz}
A.~Windisch, M.~Q. Huber, and R.~Alkofer,
  \href{http://dx.doi.org/10.1103/PhysRevD.87.065005}{{\em Phys. Rev.} {\bf
  D87} (2013) no.~6, 065005},
\href{http://arxiv.org/abs/1212.2175}{{\tt arXiv:1212.2175 [hep-ph]}}.

\bibitem{Weil:2017knt}
E.~Weil, G.~Eichmann, C.~S. Fischer, and R.~Williams,
  \href{http://dx.doi.org/10.1103/PhysRevD.96.014021}{{\em Phys. Rev.} {\bf
  D96} (2017) no.~1, 014021},
\href{http://arxiv.org/abs/1704.06046}{{\tt arXiv:1704.06046 [hep-ph]}}.

\bibitem{Pawlowski:2017gxj}
J.~M. Pawlowski, N.~Strodthoff, and N.~Wink,
  \href{http://dx.doi.org/10.1103/PhysRevD.98.074008}{{\em Phys. Rev.} {\bf
  D98} (2018) no.~7, 074008},
\href{http://arxiv.org/abs/1711.07444}{{\tt arXiv:1711.07444 [hep-th]}}.

\bibitem{Williams:2018adr}
R.~Williams, \href{http://dx.doi.org/10.1016/j.physletb.2019.134943}{{\em Phys.
  Lett.} {\bf B798} (2019)  134943},
\href{http://arxiv.org/abs/1804.11161}{{\tt arXiv:1804.11161 [hep-ph]}}.

\bibitem{Eichmann:2019dts}
G.~Eichmann, P.~Duarte, M.~T. Peña, and A.~Stadler,
  \href{http://dx.doi.org/10.1103/PhysRevD.100.094001}{{\em Phys. Rev.} {\bf
  D100} (2019) no.~9, 094001},
\href{http://arxiv.org/abs/1907.05402}{{\tt arXiv:1907.05402 [hep-ph]}}.

\bibitem{Miramontes:2019mco}
A.~S. Miramontes and H.~Sanchis-Alepuz,
  \href{http://dx.doi.org/10.1140/epja/i2019-12847-6}{{\em Eur. Phys. J.} {\bf
  A55} (2019) no.~10, 170},
\href{http://arxiv.org/abs/1906.06227}{{\tt arXiv:1906.06227 [hep-ph]}}.

\bibitem{Alkofer:2003jj}
R.~Alkofer, W.~Detmold, C.~S. Fischer, and P.~Maris,
  \href{http://dx.doi.org/10.1103/PhysRevD.70.014014}{{\em Phys. Rev.} {\bf
  D70} (2004)  014014},
\href{http://arxiv.org/abs/hep-ph/0309077}{{\tt arXiv:hep-ph/0309077
  [hep-ph]}}.

\bibitem{Windisch:2016iud}
A.~Windisch, \href{http://dx.doi.org/10.1103/PhysRevC.95.045204}{{\em Phys.
  Rev.} {\bf C95} (2017) no.~4, 045204},
\href{http://arxiv.org/abs/1612.06002}{{\tt arXiv:1612.06002 [hep-ph]}}.

\bibitem{DRL}
V.~Fran{\c{c}}ois{-}Lavet, P.~Henderson, R.~Islam, M.~G. Bellemare, and
  J.~Pineau, {\em CoRR} {\bf abs/1811.12560} (2018)  ,
  \href{http://arxiv.org/abs/1811.12560}{{\tt arXiv:1811.12560}}.
  \url{http://arxiv.org/abs/1811.12560}.

\bibitem{Sutton:2018aa}
R.~S. Sutton and A.~G. Barto, {\em Reinforcement Learning: An Introduction}.
  MIT Press, 2018.

\bibitem{Baulieu:2009ha}
L.~Baulieu, D.~Dudal, M.~S. Guimaraes, M.~Q. Huber, S.~P. Sorella,
  N.~Vandersickel, and D.~Zwanziger,
  \href{http://dx.doi.org/10.1103/PhysRevD.82.025021}{{\em Phys. Rev.} {\bf
  D82} (2010)  025021},
\href{http://arxiv.org/abs/0912.5153}{{\tt arXiv:0912.5153 [hep-th]}}.

\bibitem{Bogoliubov:1957gp}
N.~N. Bogoliubov and O.~S. Parasiuk,
\href{http://dx.doi.org/10.1007/BF02392399}{{\em Acta Math.} {\bf 97} (1957)
  227--266}.

\bibitem{Bogolyubov:1980nc}
N.~N. Bogolyubov and D.~V. Shirkov,
{\em Intersci. Monogr. Phys. Astron.} {\bf 3} (1959)  1--720.

\bibitem{Hepp:1966eg}
K.~Hepp,
\href{http://dx.doi.org/10.1007/BF01773358}{{\em Commun. Math. Phys.} {\bf 2}
  (1966)  301--326}.

\bibitem{Zimmermann:1969jj}
W.~Zimmermann, \href{http://dx.doi.org/10.1007/BF01645676}{{\em Commun. Math.
  Phys.} {\bf 15} (1969)  208--234}.
[Lect. Notes Phys.558,217(2000)].

\bibitem{Mnih:2015aa}
V.~Mnih, K.~Kavukcuoglu, D.~Silver, A.~Rusu, J.~Veness, M.~Bellemare,
  A.~Graves, M.~Riedmiller, A.~Fidjeland, G.~Ostrovski, S.~Petersen,
  C.~Beattie, A.~Sadik, I.~Antonoglou, H.~King, D.~Kumaran, D.~Wierstra,
  S.~Legg, and D.~Hassabis, \href{http://dx.doi.org/10.1038/nature14236}{{\em
  Nature} {\bf 518} (2015)  529--33}.

\bibitem{Sutton:1999aa}
R.~S. Sutton, D.~McAllester, S.~Singh, and Y.~Mansour, ``Policy gradient
  methods for reinforcement learning with function approximation,'' in {\em
  Proceedings of the 12th International Conference on Neural Information
  Processing Systems}, NIPS'99, pp.~1057--1063.
\newblock MIT Press, Cambridge, MA, USA, 1999.
\newblock \url{http://dl.acm.org/citation.cfm?id=3009657.3009806}.

\bibitem{Schulman:2015aa}
J.~Schulman, S.~Levine, P.~Abbeel, M.~Jordan, and P.~Moritz, ``Trust region
  policy optimization,'' in {\em Proceedings of the 32nd International
  Conference on Machine Learning}, F.~Bach and D.~Blei, eds., vol.~37 of {\em
  Proceedings of Machine Learning Research}, pp.~1889--1897.
\newblock PMLR, Lille, France, 07--09 Jul, 2015.
\newblock \href{http://arxiv.org/abs/1502.05477}{{\tt 1502.05477}}.
\newblock \url{http://proceedings.mlr.press/v37/schulman15.html}.

\bibitem{Schulman:2017aa}
J.~Schulman, F.~Wolski, P.~Dhariwal, A.~Radford, and O.~Klimov, {\em CoRR} {\bf
  abs/1707.06347} (2017)  , \href{http://arxiv.org/abs/1707.06347}{{\tt
  arXiv:1707.06347}}. \url{http://arxiv.org/abs/1707.06347}.

\bibitem{Kingma:2014aa}
D.~P. Kingma and J.~Ba, ``{Adam: A Method for Stochastic Optimization},'' in
  {\em {3rd International Conference on Learning Representations, {ICLR} 2015,
  San Diego, CA, USA, May 7-9, 2015, Conference Track Proceedings}}.
\newblock 2014.
\newblock \href{http://arxiv.org/abs/1412.6980}{{\tt arXiv:1412.6980 [cs]}}.

\bibitem{Paszke:2019aa}
A.~Paszke, S.~Gross, F.~Massa, A.~Lerer, J.~Bradbury, G.~Chanan, T.~Killeen,
  Z.~Lin, N.~Gimelshein, L.~Antiga, A.~Desmaison, A.~Kopf, E.~Yang, Z.~DeVito,
  M.~Raison, A.~Tejani, S.~Chilamkurthy, B.~Steiner, L.~Fang, J.~Bai, and
  S.~Chintala, in {\em Advances in Neural Information Processing Systems 32},
  pp.~8024--8035.
\newblock Curran Associates, Inc., 2019.

\bibitem{Jaderberg:2016aa}
M.~Jaderberg, V.~Mnih, W.~M. Czarnecki, T.~Schaul, J.~Z. Leibo, D.~Silver, and
  K.~Kavukcuoglu, {\em CoRR} {\bf abs/1611.05397} (2016)  ,
  \href{http://arxiv.org/abs/1611.05397}{{\tt arXiv:1611.05397}}.
  \url{http://arxiv.org/abs/1611.05397}.

\bibitem{Andrychowicz:2017aa}
M.~Andrychowicz, F.~Wolski, A.~Ray, J.~Schneider, R.~Fong, P.~Welinder,
  B.~McGrew, J.~Tobin, P.~Abbeel, and W.~Zaremba, {\em CoRR} {\bf
  abs/1707.01495} (2017)  , \href{http://arxiv.org/abs/1707.01495}{{\tt
  arXiv:1707.01495}}. \url{http://arxiv.org/abs/1707.01495}.

\bibitem{Pathak:2017aa}
D.~Pathak, P.~Agrawal, A.~A. Efros, and T.~Darrell, ``Curiosity-driven
  exploration by self-supervised prediction,'' in {\em Proceedings of the 34th
  International Conference on Machine Learning}, D.~Precup and Y.~W. Teh, eds.,
  vol.~70 of {\em Proceedings of Machine Learning Research}, pp.~2778--2787.
\newblock PMLR, International Convention Centre, Sydney, Australia, 06--11 Aug,
  2017.
\newblock \url{http://proceedings.mlr.press/v70/pathak17a.html}.

\bibitem{Alshiekh:2018aa}
M.~Alshiekh, R.~Bloem, R.~Ehlers, B.~K{\"o}nighofer, S.~Niekum, and U.~Topcu,
  ``Safe reinforcement learning via shielding,'' in {\em Proceedings of the
  Thirty-Second AAAI Conference on Artificial Intelligence, (AAAI-18), the 30th
  innovative Applications of Artificial Intelligence (IAAI-18), and the 8th
  AAAI Symposium on Educational Advances in Artificial Intelligence (EAAI-18),
  New Orleans, Louisiana, USA, February 2-7, 2018}, pp.~2669--2678.
\newblock 2018.

\end{thebibliography}\endgroup

\end{document}